\documentclass[review]{elsarticle}
\usepackage{amsmath,amssymb,amsfonts}
\usepackage{bm}
\usepackage{extarrows}
\usepackage{algorithmic}
\usepackage{algorithm}
\usepackage{subfigure}
\usepackage{bm}
\usepackage{caption}
\usepackage{graphicx}
\usepackage{color}
\usepackage{colortbl}
\journal{Knowledge-based systems}
\begin{document}

\begin{frontmatter}

\title{Disentangled Variational Auto-encoder Enhanced by Counterfactual Data for Debiasing Recommendation}
\tnotetext[mytitlenote]{}

%% Group authors per affiliation:
%\author{Elsevier\fnref{myfootnote}}
%\address{Radarweg 29, Amsterdam}
%\fntext[myfootnote]{Since 1880.}

%% or include affiliations in footnotes:
\author[mymainaddress]{Yupu Guo}
\author[mymainaddress]{Fei Cai\corref{mycorrespondingauthor}}
\cortext[mycorrespondingauthor]{Corresponding author}
\ead{18686073813@163.com}
\author[mymainaddress]{Xin Zhang}
\author[mymainaddress]{Jianming Zheng}
\author[mymainaddress]{Honghui Chen}

\address[mymainaddress]{Science and Technology on Information Systems Engineering Laboratory, National University of Defense Technology, Changsha 410073, China}
%\address[mysecondaryaddress]{National University of Defense Technology, Changsha 410073, China}

\begin{abstract}
Recommender system always suffers from various recommendation biases, seriously hindering its development.
In this light, a series of debias methods have been proposed in the recommender system, especially for two most common biases, i.e., popularity bias and amplified subjective bias.
However, existing debias methods usually concentrate on correcting a single bias.
Such single-functionality debiases neglect the bias-coupling issue in which the recommended items are collectively attributed to multiple biases.
Besides, previous work cannot tackle the lacking supervised signals brought by sparse data, yet which has become a commonplace in the recommender system.
In this work, we introduce a disentangled debias variational auto-encoder framework(DB-VAE) to address the single-functionality issue as well as a counterfactual data enhancement method to mitigate the adverse effect due to the data sparsity.
In specific, DB-VAE first extracts two types of extreme items only affected by a single bias based on the collier theory, which are respectively employed to learn the latent representation of corresponding biases, thereby realizing the bias decoupling.
In this way, the exact unbiased user representation can be learned by these decoupled bias representations.
%%
%%We first use collier theory in casual inference to find out two types of extreme items which may be affected by popularity bias and amplified subjective bias respectively. Using these extreme items as debias supervisory signals, our proposed DB-VAE can better learn an exact representation of a particular user and disentangle the complex relationship between the two biases.
%Referring to Pearl's 'abduction-action-prediction' three-step method, we design a counterfactual data generation method.
Furthermore, the data generation module employs Pearl's framework to produce massive counterfactual data, making up the lacking supervised signals due to the sparse data.
Extensive experiments on three real-world datasets demonstrate the effectiveness of our proposed model. Besides, the counterfactual data can further improve DB-VAE, especially on the dataset with low sparsity.
\end{abstract}

\begin{keyword}
Recommender systems\sep Debias\sep data sparsity.
\end{keyword}

\end{frontmatter}

%\linenumbers

%!TEX root = ./DBVAE.tex
\section{Introduction}
Recommender system(RS) {helps users find new items of personal preference from huge amounts of data}, which is extensively employed in a myriad of on-the-fly applications, such as e-commerce\citep{zhang2022price,du2021modeling}, social networks\citep{nie2019multimodal} and advertisement\citep{kingma2014adam}.
In realistic applications, the recommender system severely suffers from some widespread biases \citep{wu2021fairrec,zhu2021popularity,zhang2021causal} (e.g., the popularity bias\citep{zhu2021popularity,wei2021model} and the amplified subjective bias\citep{wang2021deconfounded}), being incapable of accurately grasping the user's personal preferences.

\begin{figure*}[b]
    \centering
    \subfigbottomskip=2pt %两行子图之间的行间距
    \subfigure[popularity bias]{\includegraphics[width=0.40\hsize,height=0.3\hsize]{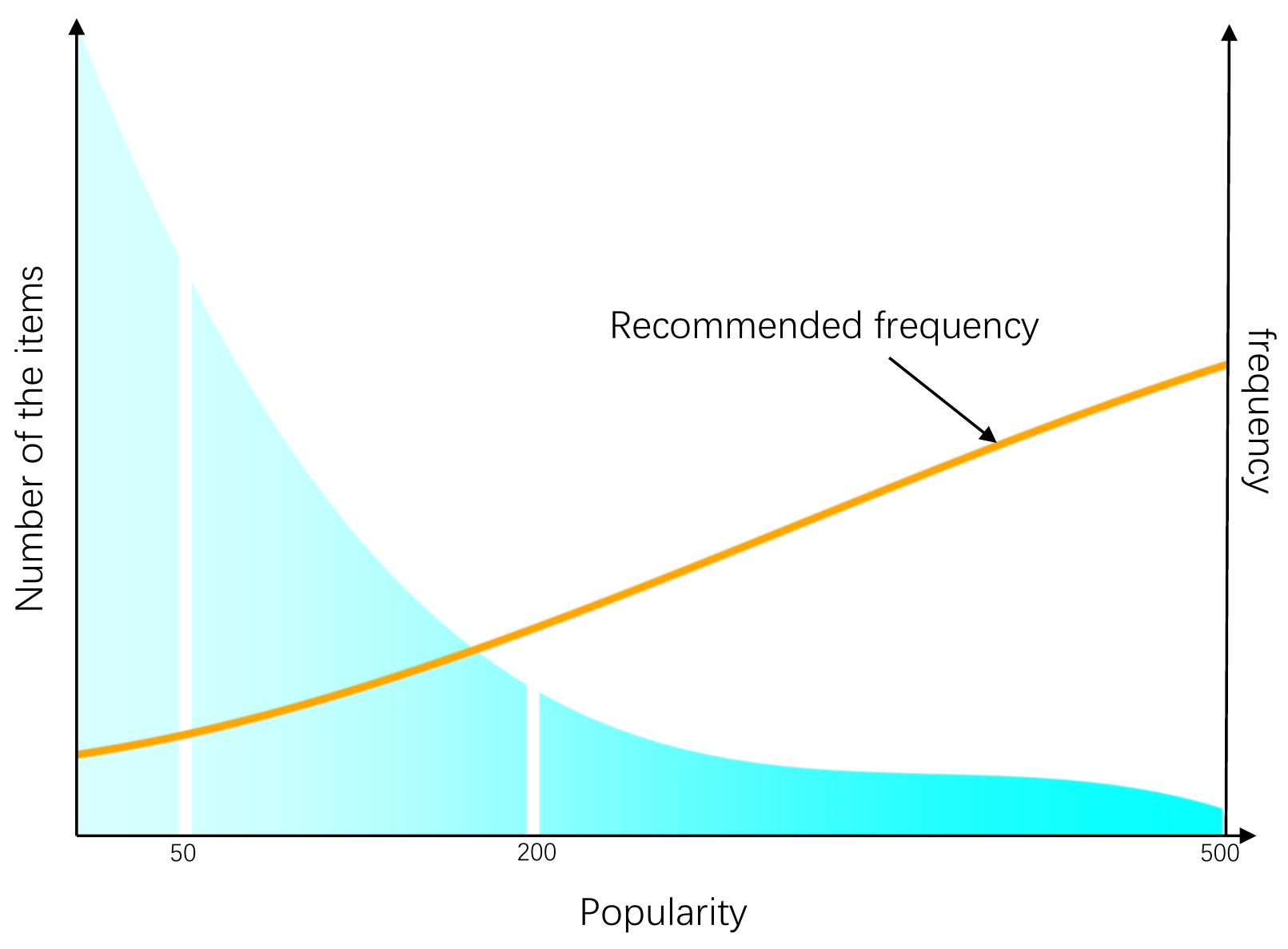}\label{Populartity}}
    \subfigure[amplified subjective bias]{\includegraphics[width=0.40\hsize,height=0.3\hsize]{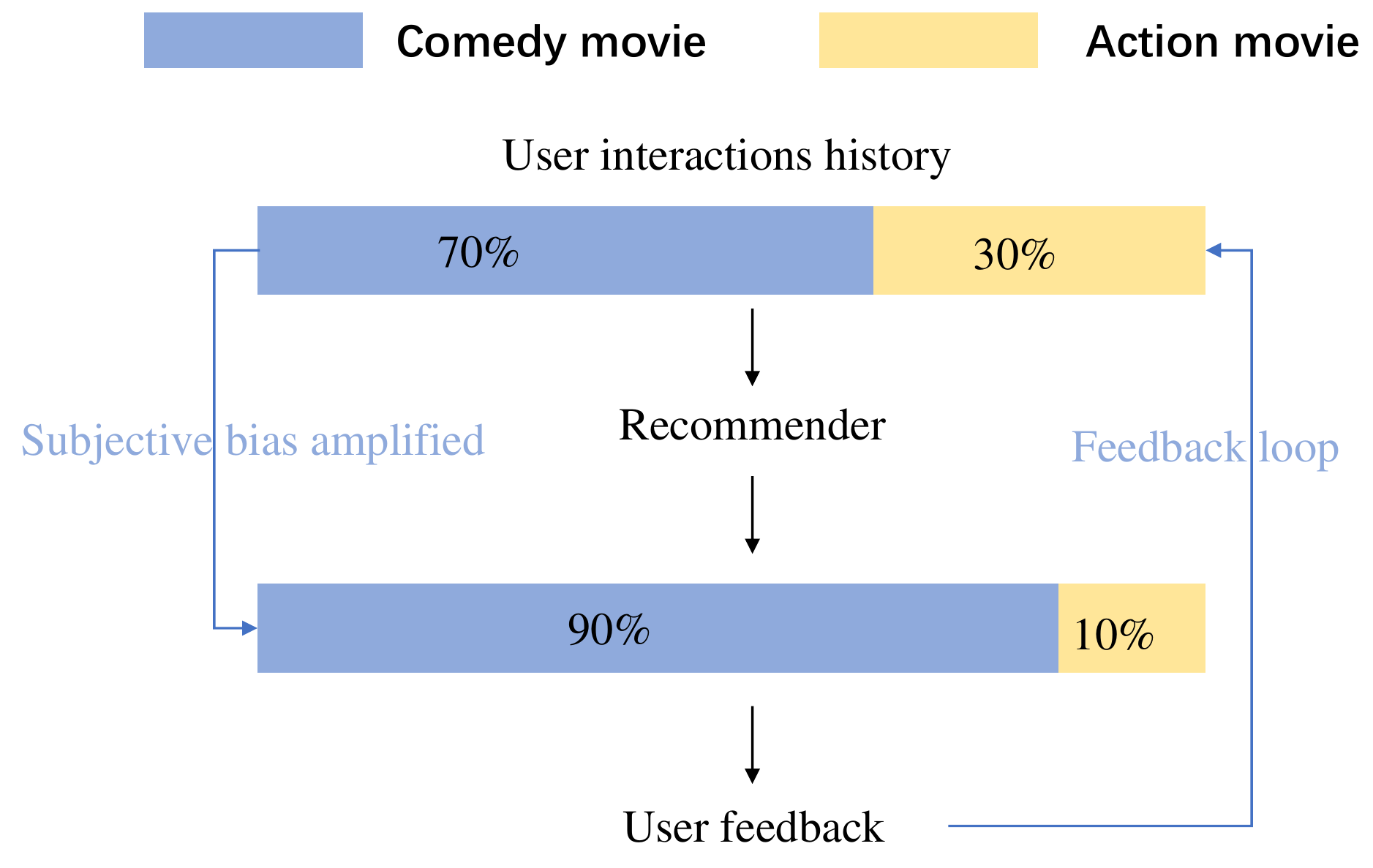}\label{abias}}
    \caption{Popularity bias and amplified subjective bias.}
    \label{Twobiases}
\end{figure*}
Fig.~\ref{Populartity}  provides an example about the popularity bias in the real-world ML-20M dataset \citep{shenbin2020recvae}.
From the left y-axis, unpopular items (less than 50 interactions) account for the vast majority of the items, yet which are rarely recommended by the traditional recommendation models (e.g., the {variational auto encoder(VAE) based model\citep{KingmaW13} } represented by the orange line at the right y-axis).
Influenced by the popularity bias, the Matthew effect\citep{wang2018quantitative} and the cold-start problem\citep{wei2021contrastive} are becoming increasingly commonplace in the recommender system.
On the other hand, the amplified subjective bias\citep{steck2018calibrated} over-recommends the history preferences of users, misinterpreting the users' current needs.
As shown in Fig.\ref{abias}, given a user's interaction history with 70\% comedies and 30\% action movies, the positive feedback loop\citep{chaney2018algorithmic} in the traditional recommenders exaggerates the user's comedy preferences, increasing to 90\% in the recommended films.
Even worse, the amplified subjective bias also triggers some issues in the recommenders (e.g., filter bubbles\citep{nguyen2014exploring} and echo chambers\citep{ge2020understanding}), causing an ever-shrinking range of the recommended items.

{Recently, a line of debias models have been proposed to alleviate these biases.
For the popularity bias, the most common strategy is item reweighting, including highlighting unpopular items\citep{liang2016modeling} or reducing the influence of popular items\citep{liang2016causal,wang2018deconfounded}.
%The main idea for alleviating popularity bias is re-weight the items according to their popularity during training, including emphasize the unpopular items\citep{liang2016modeling} or downweight the influence from popular items\citep{liang2016causal,wang2018deconfounded}.
While for the amplified subjective bias, existing models generally concentrate on adjusting the recommendation distribution from different aspects, including fairness\citep{morik2020controlling,singh2018fairness}, diversity\citep{chandar2013preference} and calibration\citep{steck2018calibrated}.
%The idea to alleviates amplified subjective bias mainly from three types: fairness\citep{morik2020controlling,singh2018fairness}, diversity\citep{chandar2013preference} and calibration\citep{steck2018calibrated}, which adjust the distribution of recommended item groups from three diffident perspectives.
%
Furthermore, some methods disentangle these biases by causal learning, e.g.,  the model-agnostic counterfactual reasoning framework (MACR)\citep{wei2021model} or the deconfounded recommendation system (DecRS) \citep{wang2021deconfounded}, of which cause-effect strategy makes the debias process more reasonable and efficient.
%In order to make more reasonable debias, several models introduce causal method\citep{wei2021model,wang2021deconfounded}.
%
%
%By exploring the popularity bias issue\citep{wei2021model} and amplified subjective bias\citep{wang2021deconfounded} from a cause-effect perspective, the debias methods become more reasonable and efficient.
}

{Albeit much progress, the above debias methods still exhibit some defects.
First is the single functionality.
Existing methods are either designed for the popularity bias, or target to the amplified subjective bias.
Such single debias methods neglect the bias-coupling probability, i.e., the items are recommended, perhaps collectively attributed to these two biases.
In this way, adjusting any sole bias is accompanied by the excessive expressiveness of the other bias.
%
%First, all the above methods mitigate the bias by limiting or adjusting confounders that may cause the bias, but such adjustment is blind without supervision signals.
Further, these bias-coupling recommended items cannot represent any pure bias, which easily causes the debias shift due to the coupled supervised signals.
%
%Second, all these debias methods are designed for a single specific bias either popularity bias or amplified subjective bias, ignoring the complex disentangled relationship between two biases.
%For example, a user interacts with an item may for two reasons, the item is popular or the item is matching his historical preference.
%If the adjustment for the popularity bias is excessive,  the matching degree will likely be the primary reason for users to interact with items, which may lead to more rampant amplified subjective bias.
%Conversely, over-limiting items that match a user's majority preference can lead to popular items dominating the recommendation list.
%
%
Besides, previous debias work doesn't consider the data-sparsity problem in the recommender system, in which quantities of users have few historical interactions (as the sparsity metric shown in Table.~\ref{table1}).
%Third, data sparsity even makes debias more difficult.
%If a user has a sufficient number of interacted items in his historical profile, there is sufficient evidence to analyze his preferences and pick out which items are affected by the bias.
%But if data is sparse, it becomes difficult to recognize users' majority preferences and bias.
%
In this light, insufficient interactions easily aggravate the difficulty of recognizing users' preferences and recommendation biases.
}

{To tackle the couple supervised issue brought by the single functionality,
%tackle the single functionality and the coupled supervised signals issues,
we propose a disentangled debias variational auto-encoder framework (DB-VAE) that decouples these two biases to help obtain a debias representation of user preferences.
In particular,
we first analyze the reason about the interaction between users and items, from which two direct causes of user click behavior are summarized: the item is popular and the item is matching the user's preference.
From this analysis, we design a casual graph about the user click behavior, helping disentangle the popularity bias and amplified subjective bias from the user's interaction profile \citep{pearl2010causal}.
%%
%Based on the colliding effect in casual theory\citep{pearl2010causal}, this causal graph helps disentangle the popularity bias and amplified subjective bias from the user's interaction profile.
By defining the data split criterion of disentangled biases, two extreme types of items only affected by one bias can be extracted to further train the debias process, thereby mitigating the adverse effect of couple supervised signals.
%Through the variational auto-encoder framework with two debias supervision signals, our proposed DB-VAE aims to learn a debias representation of user preferences.
%
In addition, to address the data sparsity issue, we further design a counterfactual data generation strategy to produce massive counterfactual data, making up the lacking supervised signals due to the sparse interactions.
%In order to alleviate the difficulties caused by data sparsity, we follow Pearl's causal inference framework\citep{peters2017elements} and design a counterfactual data generation strategy to help train the DB-VAE.}
In this generation process, a Pearl's causal inference framework\citep{peters2017elements} is employed} {to help answer the questions about how will user perform if he or she is not affected by biases. Following the 'Abduction-Action-Prediction' process, we answer the counterfactual questions and generate the corresponding counterfactual data, making up the lacking supervised signals}.

{We evaluate DB-VAE on three real-world RS datasets to verify its effectiveness.
Extensive experimental results show that DB-VAE outperforms state-of-the-art baselines with average 5\% improvements in terms of $NDCG$ and $Recall$.
Furthermore, generating counterfactual data can further enhance DB-VAE, especially on sparser datasets.}
{Overall, the contributions of this work can be summarized as the following three folds:}

\begin{enumerate}
\item {presenting a disentangled debias variational auto-encoder which simultaneously eliminates the popularity bias and the amplified subjective bias by two extreme types of items;}
\item {a counterfactual data generation method to make up the lacking supervised signals due to the data sparsity;}
\item {evaluations on three real-world datasets demonstrating the effectiveness and rationality of our proposed models.}
\end{enumerate}
%!TEX root = ./DBVAE.tex
\section{Related work}

In this section, we review existing recommendation debias work, mainly concentrating on the popularity bias and the amplified subjective bias.

\subsection{Popularity Bias in Recommendation}
In the recommender system, it is a common problem that users' feedback is easily influenced by popular items.
Recently, eliminating the popularity bias has received much attention in the recommendation research.
Specifically, methods targeting to de-bias the popularity bias can be roughly divided into the following three types.
The first line of the debias research aims at reweighting the popular items during training.
For instance, \citet{rosenbaum1983central} firstly propose an inverse propensity weighting (IPW) to adjust the importance of items according to their popularities.
Inspired by this weighting strategy, \citet{liang2016causal} propose an improved method to impose lower weights on popular items.
The second line of the debias research takes advantage of the ranking adjustment to calibrate the popularity bias.
For example, \citet{abdollahpouri2017controlling} introduce a regularization-based method to promote the rank of unpopular items.
In addition, \citet{abdollahpouri2019managing} adopt a re-ranking strategy to adjust the output rank of the recommender system.
The third line of the debias research employs the casual theory to alleviate the popularity bias, including the causal representation learning\citep{liu2020general,zheng2021disentangling} and the causal adjustment\citep{zhang2021causal,wei2021model}.

{Unlike the existing work, we argue that popularity bias should not be targeted singly because there is a complex coupled relationship between the popularity bias and the other biases.
}
{Hence, we introduce two extreme types of items to further train the debias process, thereby overcoming the blindness of debias brought by the couple biases. }
%With the help of casual theory, we extracted extreme interactions from user historical behaviour as debias supervisory signals.}

\subsection{Amplified subjective bias}
During training, the positive feedback loop forces the recommender system to amplify the user's historical preferences \citep{chaney2018algorithmic}.
To correct such an amplified subjective bias, existing debias methods mainly explore from the following three aspects, i.e., fairness, diversity and causal recommendation. For the fairness in RS, \citet{biega2018equity} propose an amortized equity of attention, which ensures that similar individuals receive the similar treatments.
And \citet{morik2020controlling} argue all user groups are supposed to be treated fairly, thereby introducing a more granular way to keep the group fairness.
Besides, \citet{steck2018calibrated} re-rank the items to ensure the distribution of the recommended item groups to be matched with the users' interaction history.
For the diversity in RS, existing work pursues the dissimilarity of the recommended items \citep{cheng2017learning,sun2020framework}, where similarity can be measured by item category and embeddings\citep{chandar2013preference,jiang2020aspect}.
Furthermore, the causal recommendation targets to find the root reason for amplified bias.
For example, \citet{wang2021deconfounded} first scrutinize the cause-effect factors for bias amplification, and then contribute an approximation operator to eliminate the amplified subjective bias by the back-door adjustment.

{However, the solely-modeling amplified subjective bias easily cause serious filter bubbles\citep{nguyen2014exploring} and echo chambers\citep{ge2020understanding}.
Hence, we consider the amplified subjective bias combined with the other bias (mainly the popularity bias) during debias process.}
{In this light, our proposed model eliminates the two types of bias simultaneously to disentangle the complex coupled relationship between these two biases.}

%!TEX root = ./DBVAE.tex
\section{Method}
%In this section, we first illustrate how to extract interacted items which may be subjected to bias from user's historical interactions.
%Then, we elaborate the proposed DBVAE model.
%Finally, the counterfactual data enhancement method is introduced.
{In this section, we first present the bias extraction in \S \ref{sec:method_bias} to construct two extreme types of items only affected by a single bias.
Then, we elaborate on the proposed DB-VAE model in \S \ref{sec:method_debias}, which helps accurately characterize the user's debias latent representation and give debias predictions for user's preferences.
Finally, we introduce a counterfactual data generation method in \S \ref{Counterfactual} to enhance DB-VAE to tack the data sparsity issue.}
\subsection{Bias extraction}
\label{sec:method_bias}

\begin{figure}[h]
	\centering
	\includegraphics[width=0.7\hsize,height=0.3\hsize]{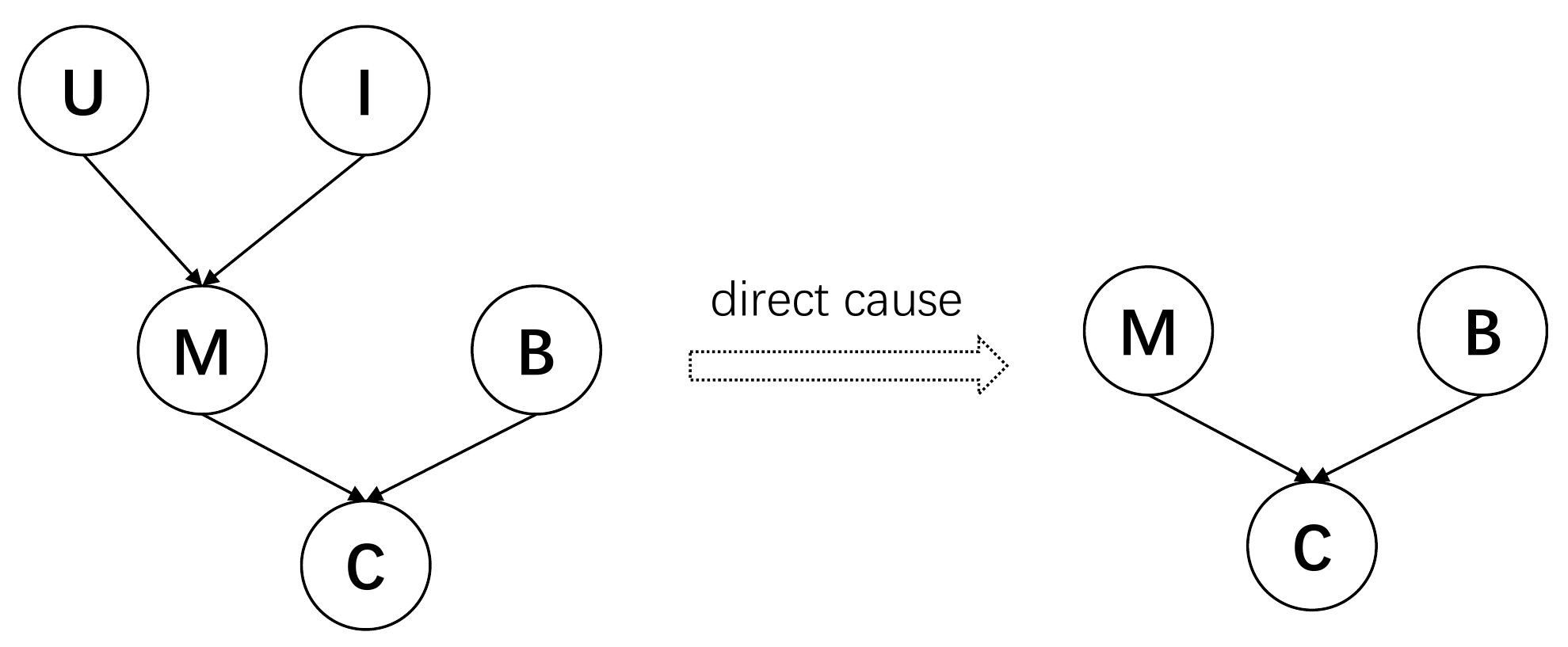}
	\caption{Causal graph on users interactions. Node C denotes the click behavior of user. Node B and Node M represent the popularity of the specific item and the matching degree between the item (Node I) and the user (Node U), respectively.}
	\label{fig1}
\end{figure}

{According to the causal theory \citep{pearl2010causal}}, we design a causal graph on users' interactions in Fig.~\ref{fig1}.
Clearly, there exist two direct causes (Node B and Node M) towards the users' click behavior (Node C), i.e., the popularity of a specific item and the matching degree between this item and the target user.
%Node C, B and M represent the users click behavior and its two causes respectively.
%
Although such two causes are independent in essence (Node B and Node M are not directly connected), their collider (Node C) makes them correlated in the expressive form of recommendation items.
%And such three nodes constitute a \emph{immorality}, in which click(C) is a \emph{collider} of popularity(B) and matching degree(M)\citep{pearl2018book,peters2017elements}.
%%
%%Generally, Node B and Node M in Fig.~\ref{fig1} are independent.
%%However, in causal theory\citep{pearl2010causal}, two causes(B and M) become correlated when we condition on the collider(C).
%Following the causal theory\citep{pearl2010causal}, such three nodes constitute a \emph{immorality}.
%
In this way, we expect to obtain the click-behavior data attributed to the sole cause.
According to the \emph{colliding effect}\citep{pearl2018book,peters2017elements}, two extreme conditions can be extracted from user interactions, which reflect the popularity bias and the amplified subjective bias, respectively.
In particular, if a user clicks on an item that hardly matches his/hers historical preferences, this click behavior is stemmed from the fact that this recommended item is very popular and the user is affected by conformity.
And vice versa, an unpopular item gets clicked owing to its compatibility with the user's personal preferences.
%This phenomena is called \emph{colliding effect}.
%According to the colliding effect, we can select two extreme types of user interaction, which may result in popularity bias and amplified subjective bias.

%In order to facilitate the comparison of popularity and matching degree with the user between different items in the interaction profile of the target user, we first define popularity score $S_m(u,i)$ and matching score $S_p(u,i)$ for every item that the target user interacts with.
To measure the item popularity and the matching degree between users and items, the popularity score $S_m(u,i)$ and the matching score $S_p(u,i)$ between the target user $u$ and the recommended item $i$ are respectively introduced, as follows:
%
%\begin{equation}
%S_p(u,i)= \frac{num(i)}{\sum_{j\in{X}}num(j)},
%\label{eq1}
%\end{equation}
%\begin{equation}
%S_m(u,i)= \frac{\textbf{d}_i\cdot\textbf{d}_u}{|\textbf{d}_u|},
%\label{eq2}
%\end{equation}
%
\begin{equation}
 	\left\{\begin{array}{ll}
 		S_p(u,i)= \frac{num(i)}{\sum_{j\in{X}}num(j)},
 		\\
 		S_m(u,i)= \frac{\textbf{d}_i\cdot\textbf{d}_u}{|\textbf{d}_u|},
 	\end{array}\right.
 	\label{similarity}
\end{equation}
where $X$ denotes the interaction profile with all items the target user $u$ clicked before and $num(i)$ is the click number of the item $i$ in the recommender system.
And the popularity score $S_p(u,i)$ is defined as the ratio between the click number of the item $i$ and the click number summation of items the user $u$ clicked before.
%Because the popularity of an item is compared between all items clicked by the same user, $num(i)$ is divided by the sum of the popularity of all items clicked by the target user $u$ to construct the popularity score $S_p(u,i)$.
%$\textbf{d}_i$ and $\textbf{d}_u$ are category vectors of item $i$ and user $u$.
Besides, $\textbf{d}_i$ and $\textbf{d}_u$ denote the category representations of item $i$ and user $u$ by the one-hot format, respectively.
In specific, given that there are five categories of movie tags in a movie dataset, including 'Comedy','Action', 'Crime', 'Adventure' and 'Thriller', the dimension of the category embeddings can be set to 5.
In this way, when a movie has 'spider man' has 'Comedy' and 'Action' tags, its category embeddings can be represented as $[1,1,0,0,0]$.
While the user category embeddings are obtained by summarizing the category embeddings of all items the user has clicked on, i.e.,
\begin{equation}
\textbf{d}_u = \Sigma_{i \in X}\textbf{d}_i,
\end{equation}

Next, we elaborate on two extreme conditions from user interactions to obtain the sole-cause items.
The first is the items that are clicked due to the conformity, which causes the popularity bias.
So we select items that are very popular but rarely match the target user as $X_p$, i.e.,
%
%items that are very popular but rarely matches the target user.
%Users clicking on these items are likely to be influenced by conformity. If there are many such items involved in the training of the recommendation system, the system may be trapped in popularity bias. s:
%
\begin{equation}
X_p = \{i|rank(S_p(u,i))<k|X|,rank(S_m(u,i))>k|X|,i\in{X}\}
\label{eq4}
\end{equation}
where $rank(S_p(u,i))$ and $rank(S_m(u,i))$ represents the ranking of the popularity score and the matching score of the item $i$ among all items clicked by the target user $u$, respectively;
and $k\in[0,1]$ denotes the debias degree in which the smaller k indicates that less items are extracted from user's interactions to construct $X_p$.

The second is the items that are clicked due to the user's interests, which amplifies the subjective bias during the training process, and probably makes the recommender system fall into information cocoons\citep{hou2021information}.
%
%are those have a good match for a user's interests, but are rarely clicked by other users, i.e. unpopular.
%These items may amplify the subjective bias during the training process and may let users fall into information cocoons\citep{hou2021information}.
Similarly, we select items that have a good match for the user's interests, yet are rarely clicked by other users as $X_m$, i.e.,
\begin{equation}
X_m = \{i|rank(S_p(u,i))>k|X|,rank(S_m(u,i))<k|X|,i\in{X}\}
\label{eq5}
\end{equation}
\subsection{Debias variational auto-encoder}
\label{sec:method_debias}

\begin{figure}[h]
	\centering
	\includegraphics[width=1\textwidth]{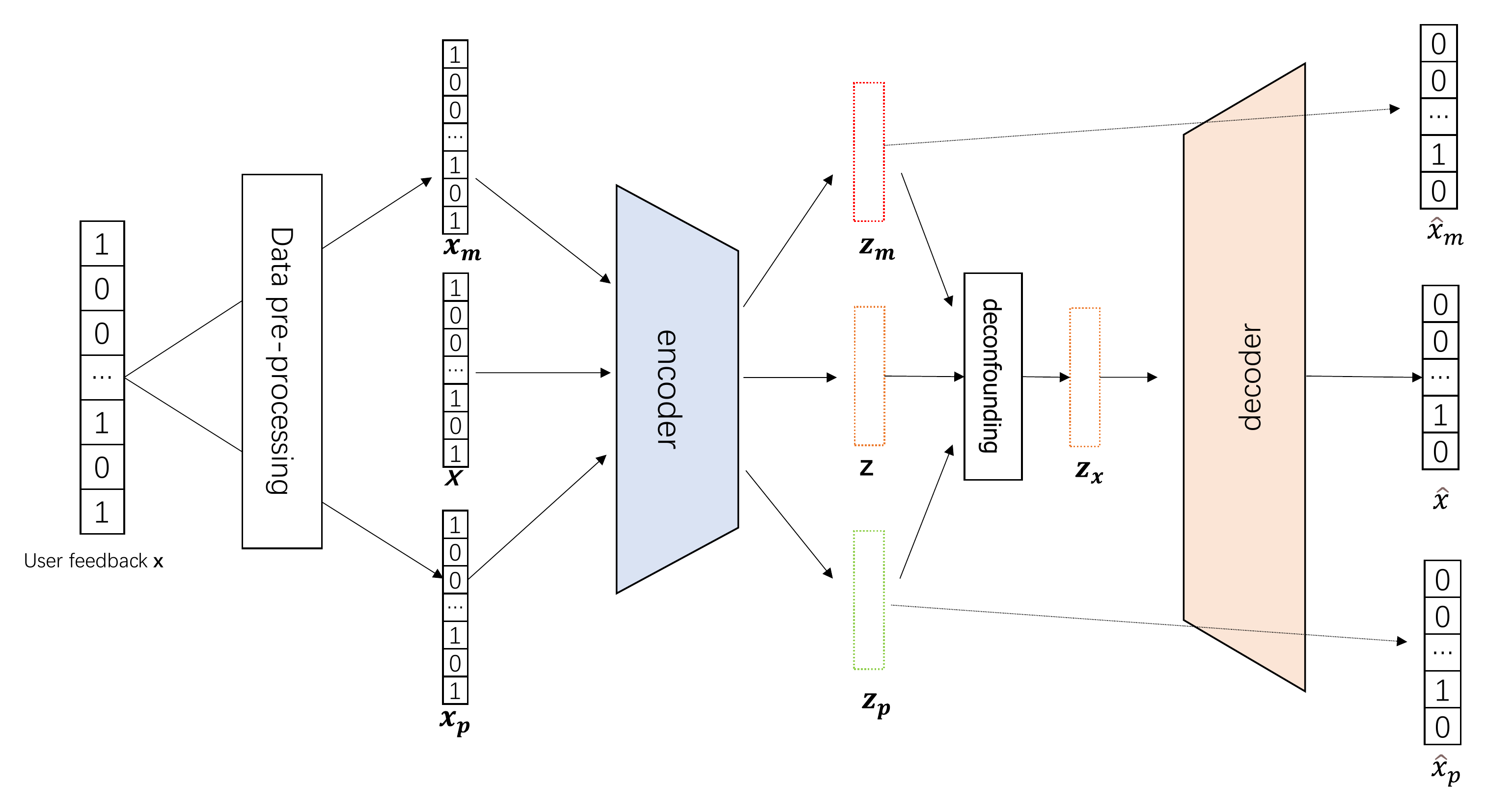}
	\caption{Framework of DB-VAE.}
	\label{fig2}
\end{figure}

Variational auto-encoder(VAE) is a deep latent variable model\citep{KingmaW13,RezendeMW14}.
By the encoder-decoder structure, VAE can effectively learn latent representation from real-world data.
Different from conventional auto-encoder(AE), VAE tries to ensure that the latent space is regular enough by introducing a distribution regularization during the training process.
In this way, VAE can alleviate the over-fitting problem in conventional AE.
In view of these merits, we select VAE as the backbone of DB-VAE in the hope that the standardized feature representation of users can be learned during the debias process.
We present the framework of DB-VAE in Fig.\ref{fig2}.

As shown in Fig.\ref{fig2}, the total interacted items of the target user $X$, items $X_p$ attributed to the popularity bias and items $X_m$ attributed to the amplified subjective bias are taken as the input of VAE.
%
%We take $X$, $X_m$ and $X_p$ as the input of VAE, which contain the total interacted items of the target user, items that may cause popularity bias in $X$ and items that may cause amplified subjective bias in $X$ respectively.
%Before being put into VAE, $X$, $X_m$ and $X_p$ are encoded as one-hot vectors $\bm{x}$, $\bm{x_m}$ and $\bm{x_p}$. If item $i$ is in $X$, then the corresponding position of $\bm{x}$ is set to 1, otherwise, the corresponding position is set to 0.
And then, we employ the one-hot format embedding to encode $X$, $X_m$ and $X_p$, obtaining the corresponding embeddings $\bm{x}$, $\bm{x_m}$ and $\bm{x_p}$.
These one-hot embeddings are further fed into the VAE encoder to generate their latent representations $\bm{z_x}$, $\bm{z_m}$ and $\bm{z_p}$, which can be formulated as follows:

\begin{equation}
 	\left\{\begin{array}{ll}
 		\bm{z_x}\sim p_\phi(\bm{z_x}|\bm{x}),
 		\\
 	\bm{z_m} \sim p_\phi(\bm{z_m}|\bm{x_m}),
 	\\
 	\bm{z_p} \sim p_\phi(\bm{z_p}|\bm{x_p}),
 	\end{array}\right.
\end{equation}
%\begin{equation}
%\bm{z_x}\sim p_\phi(\bm{z_x}|\bm{x}),
%\end{equation}
%%
%\begin{equation}
%\bm{z_m} \sim p_\phi(\bm{z_m}|\bm{x_m}),
%\end{equation}
%%
%\begin{equation}
%\bm{z_p} \sim p_\phi(\bm{z_p}|\bm{x_p})
%\end{equation}
%
where $\phi$ denotes the parameter set of the VAE encoder.
Here, we select a 5-layer neural network as the VAEencoder.
Furthermore, $\bm{z_p}$ and $\bm{z_m}$ can be treated as the user representations influence by the popularity bias and the amplified subjective bias, respectively.
To reconstruct a user latent representation vector that is not affected by these two biases, we can obtain the unbiased user representation by subtracting $\bm{z_p}$ and $\bm{z_m}$ from $\bm{z_x}$, i.e.,
\begin{equation}
\bm{z}=\bm{z_x}-\bm{z_m}-\bm{z_p}
\label{eqsub}
\end{equation}

Different from the conventional auto-encoder, VAE assumes that the latent representation is subject to a prior standard normal distribution, from which a random sampling can be decoded to real-world data $\bm{x}$.
However, the random sampling operation is not differentiable, which makes the model untrainable.
In this light, we adopt the standard normal distribution $\mathcal{N}(\mu,\sigma ^2)$ as the posterior distribution generated by the encoder, and make the output of VAE approach this distribution by the KL divergence constraint during the model training. Formally, the output of the VAE encoder is subject to:
\begin{equation}
p_\phi(\bm{z_x}|\bm{x})= \mathcal{N}(\bm{{\mu}_x},\bm{{\sigma}_x}^2),
\end{equation}
where
\begin{equation}
 	\left\{\begin{array}{ll}
 		\bm{{\mu}_x}= g_1(f_{\phi}(\bm{x})),
 		\\
 	log\bm{{\sigma}_x}^2=g_2(f_{\phi}(\bm{x})),
 	\\
 	\end{array}\right.
 	\label{eqencoder}
\end{equation}
%\begin{equation}
%\bm{{\mu}_x}= g_1(f_{\phi}(\bm{x}))
%\end{equation}
%\begin{equation}
%log\bm{{\sigma}_x}^2=g_2(f_{\phi}(\bm{x}))
%\label{eq12}
%\end{equation}
in which $f_{\phi}()$ is a multilayer perceptron in the VAE encoder, $g_1()$ and $g_2()$ are two different fully connected layers of the VAE encoder.
Analogously, the poster distributions $p_{\phi}(\bm{z_m}|\bm{x_m})$ and $p_{\phi}(\bm{z_p}|\bm{x_p})$ are generated in the same way.
VAE takes the representation vectors $\bm{z_x}$, $\bm{z_m}$ and $\bm{z_m}$ as samples from the generated distribution $\mathcal{N}(\bm{{\mu}_x},\bm{{\sigma}_x}^2)$, $\mathcal{N}(\bm{{\mu}_m},\bm{{\sigma}_m}^2)$ and $\mathcal{N}(\bm{{\mu}_p},\bm{{\sigma}_p}^2)$.

In order to ensure this operation is derivable, a re-parameterized way is adopted, which can be formulated as follows:
\begin{equation}
 	\left\{\begin{array}{ll}
 		\bm{z_x} = \bm{\mu_x} + \bm{\epsilon}{\odot}\bm{\sigma_x},
 		\\
 	\bm{z_m} = \bm{\mu_m} + \bm{\epsilon}{\odot}\bm{\sigma_m},
 	\\
 	\bm{z_p} = \bm{\mu_p} + \bm{\epsilon}{\odot}\bm{\sigma_p},
 	\end{array}\right.
\end{equation}
%
%\begin{equation}
%\bm{z_x} = \bm{\mu_x} + \bm{\epsilon}{\odot}\bm{\sigma_x}
%\end{equation}
%%
%\begin{equation}
%\bm{z_m} = \bm{\mu_m} + \bm{\epsilon}{\odot}\bm{\sigma_m}
%\end{equation}
%%
%\begin{equation}
%\bm{z_p} = \bm{\mu_p} + \bm{\epsilon}{\odot}\bm{\sigma_p}
%\end{equation}
where $\bm{\epsilon}$ is a sampled vector from multivariate standard normal distribution $\mathcal{N}(0,\bm{I})$ and $\odot$ is element-wise product operation.
In this way, the sampling $\mathcal{N}(\bm{{\mu}_x},\bm{{\sigma}_x}^2)$ is transformed to the sampling $\mathcal{N}(0,\bm{I})$.
% and then obtain the result of sampling from $\mathcal{N}(\bm{{\mu}_x},\bm{{\sigma}_x}^2)$ by parameter transformation.
Hence, the operation 'sampling' is not involved in the gradient descent, but in the result of sampling, making the whole model trainable.
Under this sampling, the debiased user latent representation $\bm{z}$ in Eq.
\ref{eqsub} can be transformed into:
\begin{equation}
\bm{z} = \bm{\mu_x} -\bm{\mu_m}-\bm{\mu_p}+ \bm{\epsilon}{\odot}(\bm{\sigma_x}-\bm{\sigma_m}-\bm{\sigma_p})
\end{equation}
With these laten representation, the final prediction score $\hat{\bm{x}}$ and the prediction scores of two biases ($\hat{\bm{x}}_p$ and $\hat{\bm{x}}_m$) can be decoded as follows:
%Then the final prediction score $\hat{\bm{x}}$ of the user can be obtained by decoding the debiased user latent representation $\bm{z}$. Also, biased prediction score $\hat{\bm{x}}_p$ and $\hat{\bm{x}}_m$ can be obtained in the same way:
%

\begin{equation}
 	\left\{\begin{array}{ll}
 	\hat{\bm{x}}\sim q_{\theta}(\hat{\bm{x}}|\bm{z}),
 		\\
 	\hat{\bm{x}}_p\sim q_{\theta}(\hat{\bm{x}}_p|\bm{z_p}),
 	\\
 	\hat{\bm{x}}_m\sim q_{\theta}(\hat{\bm{x}}_m|\bm{z_m}),
 	\end{array}\right.
\end{equation}
%\begin{equation}
%\hat{\bm{x}}\sim q_{\theta}(\hat{\bm{x}}|\bm{z})
%\end{equation}
%\begin{equation}
%\hat{\bm{x}}_p\sim q_{\theta}(\hat{\bm{x}}_p|\bm{z_p})
%\end{equation}\
%\begin{equation}
%\hat{\bm{x}}_m\sim q_{\theta}(\hat{\bm{x}}_m|\bm{z_m})
%\end{equation}
%
where $\theta$ denotes the parameter set of the decoder.

Following the VAE training, the encoder parameter set $\phi$ and the decoder parameter set $\theta$ can be learned by maximizing the evidence lower bound (ELBO)\citep{blei2017variational}:

\begin{equation}
\mathcal{L}^{ELBO}(\hat{\bm{x}};\theta,\phi) = E_{p_{\phi}(\bm{z_x}|\bm{x},\bm{x_p},\bm{x_m})}
[logq_{\theta}(\hat{\bm{x}}|\bm{z_x})-KL(p_{\phi}(\bm{z_x}|\bm{x},\bm{x_p},\bm{x_m})||p(\bm{z_x}))]
\end{equation}
where the first half of this formula, $E_{p_{\phi}}[logq_{\theta}(\hat{\bm{x}}|\bm{z_x})]$ narrows the gap between $\bm{x}$ and $\hat{\bm{x}}$, and hence the cross entropy is employed to achieve this goal;
the second part $KL(p_{\phi}(\bm{z_x}|\bm{x},\bm{x_p},\bm{x_m})||p(\bm{z_x}))$ represents the KL divergence between the posterior distribution $p_{\phi}(\bm{z_x}|\bm{x},\bm{x_p},\bm{x_m})$ and the prior distribution $p(\bm{z_x})$ with the assumption $p(\bm{z_x})\sim \mathcal{N}(0,\bm{I})$.
By optimizing the negative KL divergence, DB-VAE can try to force that the debiased user feature representation $\bm{z_x}$ adheres to the standard normal distribution.

%With the purpose to keep the model's ability to accurately represent biased data and achieve better debias,
To overcome the issue of coupled supervised signals in the previous VAE training, we employ the bias-extracted data $X_p$ and $X_m$ to further train the VAE decoder, ensuring the accurate generation of $\bm{z_m}$ and $\bm{z_p}$:

\begin{equation}
 	\left\{\begin{array}{ll}
 	\mathcal{L}^s(\hat{\bm{x_p}};\theta,\phi)=E_{p_{\phi (\bm{z_p}|\bm{x_p})}}[logq_{\theta}(\bm{\hat{x}_p}|\bm{z_p)}],
 		\\
 	\mathcal{L}^s(\hat{\bm{x_m}};\theta,\phi)=E_{p_{\phi (\bm{z_m}|\bm{x_m})}}[logq_{\theta}(\bm{\hat{x}_m}|\bm{z_m)}],\\
 	\bm{x_p}, \bm{x_m} = one-hot (X_p), one-hot (X_m),
 	\end{array}\right.
\end{equation}
where $one-hot(\cdot)$ is the one-hot embedding.
%
%\begin{equation}
%\mathcal{L}(\hat{\bm{x_p}};\theta,\phi)=E_{p_{\phi (\bm{z_p}|\bm{x_p})}}[logq_{\theta}(\bm{\hat{x}_p}|\bm{z_p)}]
%\end{equation}
%\begin{equation}
%\mathcal{L}(\hat{\bm{x_m}};\theta,\phi)=E_{p_{\phi (\bm{z_m}|\bm{x_m})}}[logq_{\theta}(\bm{\hat{x}_m}|\bm{z_m)}]
%\end{equation}
%
All in all, the final loss can be jointly connected as:
\begin{equation}
\mathcal{L}=\mathcal{L}^{ELBO}(\hat{\bm{x}};\theta,\phi)+\omega_{m}\mathcal{L}^s({\bm{\hat{x}_m}};\theta,\phi)
+\omega_{p}\mathcal{L}^s({\bm{\hat{x}_p};\theta,\phi)},
\end{equation}
where $\omega_{m}$ and $\omega_{p}$ are weight parameters.

\subsection{Counterfactual data generation}
\label{Counterfactual}

Intuitively, the direct way to mitigate the adverse effect of sparse data is to increase the data volume.
To this end, we employ Pearl's causal inference framework\citep{peters2017elements} to generate counterfactual data, in which this inference framework contains a three-layer causal hierarchy, including abduction, intervention and prediction.

\begin{figure}[h]
	\centering
	\includegraphics[width=0.5\hsize,height=0.3\hsize]{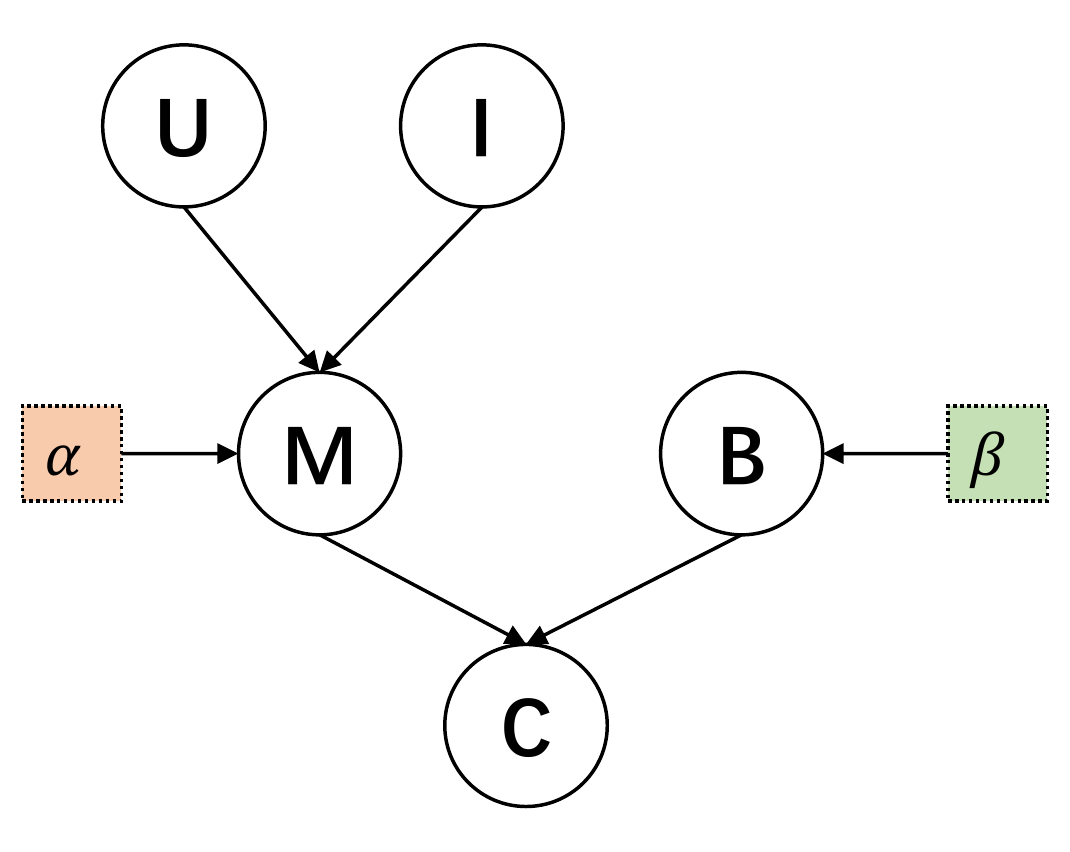}
	\caption{Casual graph with exogenous variables.}
	\label{fig3}
\end{figure}

Following this hierarchy, we first construct a basic model $\mathbf{F}$ abided by the causal graph proposed
in Fig.\ref{fig1}.
Different from the causal graph, two exogenous variables $\alpha$ and $\beta$ are introduced to ensure the counterfactual prediction works.
In specific, exogenous variables $\alpha$ and  $\beta$ describe uncertainties that affect the matching degree in node $M$ and the popularity attribute in node $B$, respectively.
For example, Santa movies are probably more popular around Christmas, revealing that the extra factor of date potentially influences the popularity.
This indicates that there exist some uncertainties affecting the data generation process.
Hence, the basic model $\mathbf{F}$ (as shown in Fig.\ref{fig3}) is defined in a stochastic manner to consider the randomness and possible noisy data.

%

%$\alpha$ is the exogenous variable of the matching degree(node $M$) while $\beta$ is the exogenous variable of the popularity(node $B$).
%

%Hence, the casual graph are transformed into Fig.\ref{fig3}.
%In a stochastic manner, the basic generation process of counterfactual data can be formulated as follows:
Formally, according to the inference process in Fig.\ref{fig3}, the basic model $\mathbf{F}$ can be formulated as follows:
\begin{equation}
\mathbf{F}:\left\{
\begin{aligned}
&M\sim{p_M(M|U,I,\alpha)}\\
&B\sim{p_B(B|\beta)} \\
&C\sim{p_C(C|M,B)}\\
&\alpha,\beta \sim{\mathcal{N}(0,I)}
\end{aligned}
\right.,
\end{equation}
%
%%The structure equation models(i.e.,$F$) are defined in a stochastic manner in order to consider randomness and possible noisy data.
%%The exogenous variables $\alpha$ and $\beta$ can be seen as uncertainties that may affect the matching degree and the popularity of items. For example, Santa movies are probably more popular around Christmas, which reveals that date is a potential factor influencing popularity. This uncertainties is latent but can affect the data generation process.
%%'Abduction' is the key step which aims to infer the posteriors of the exogenous variables $\alpha$ and $\beta$. Before 'Abduction', we need to learn a specific function $F$.
%
where the distributions $p_M$, $p_B$ and $p_C$ can be detailed as:
\begin{equation}
 	\left\{\begin{array}{ll}
 	M(u,i)= \frac{exp(\mathbf{E}_u^T \mathbf{E}_i + w_i^M \cdot S_m(u,i) \cdot \alpha_i)}{\sum_{j}^{|\emph{I}|}exp(\mathbf{E}_u^T \mathbf{E}_j + w_j^M \cdot S_m(u,j) \cdot \alpha_j)},
 		\\
 	B(i)= \frac{exp(w_i^B\cdot S_p(u,i) \cdot \beta_i)}{\sum_{j}^{|\emph{I}|}exp(w_j^B \cdot S_p(u,j) \cdot \beta_j)},
 	\\
 	C(u,i)= \frac{exp(M(u,i)+B(i))}{\sum_{j}^{|\emph{I}|}exp(M(u,j)+B(j))},
 	\end{array}\right.
 	\label{generation}
\end{equation}
%\begin{equation}
%M(u,i)= \frac{exp(\mathbf{E}_u^T \mathbf{E}_i + w_i^M \cdot S_m(u,i) \cdot \alpha_i)}{\sum_{j}^{|\emph{I}|}exp(\mathbf{E}_u^T \mathbf{E}_j + w_j^M \cdot S_m(u,j) \cdot \alpha_j)}
%\label{eq18}
%\end{equation}
%\begin{equation}
%B(i)= \frac{exp(w_i^B\cdot S_p(u,i) \cdot \beta_i)}{\sum_{j}^{|\emph{I}|}exp(w_j^B \cdot S_p(u,j) \cdot \beta_j)}
%\label{eq19}
%\end{equation}
%\begin{equation}
%C(u,i)= \frac{exp(M(u,i)+B(i))}{\sum_{j}^{|\emph{I}|}exp(M(u,j)+B(j))}
%\label{eq20}
%\end{equation}
%
{where $\mathbf{E}_u$ and $\mathbf{E}_i$ are the embeddings of user $u$ and item $i$ respectively;
${w}^B_i$ and ${w}^M_i$ are weighting parameters which can be learned during the training process,} and $S_m(u,i)$ and $S_p(u,i)$ can be referred to Eq.\eqref{similarity}.
%, indicating the matching degree between user $u$ and item $i$ and the popularity score of item $i$ in $u$'s interaction profile.

With the observed dataset $\textit{O}=\cup_{i=1}^{|I|}\cup_{u=1}^{|U|}\{(x(u,i))\}$, the basic model $F$ can be learned by the cross entropy loss on the click prediction module $C(u,i)$, i.e.,
\begin{equation}
\mathcal{L}_F = \sum_{u}^{|U|} \sum_{i}^{|I|}[x(u,i)logC(u,i)]
\end{equation}
where $x(u,i)$ is the ground truth indicating whether the user $u$ interacted with the item $i$.
%By optimizing the entropy loss, a specific $F$ is learned from the observed training data.
%During the learning process, we assume that
To ensure the randomness during th model training, $\mathbf{\alpha}$ and $\mathbf{\beta}$ are subject to the multivariate standard normal distribution $\mathcal{N}(0,I)$.
Since the distribution of the observed data is highly influenced by $\mathbf{\alpha}$ and $\mathbf{\beta}$, the posterior distributions in different datasets are diverse.
In this way, once we have learned $F$, we can follow the Pearl's abduction-action-prediction framework\citep{glymour2016causal} to generate counterfactual data, further strengthening the DB-VAE training.

In particular, the main objective of the abduction process is to estimate the posterior of $\mathbf{\alpha}$ and $\mathbf{\beta}$ from the observed dataset $\textit{O}$.
Taking $\mathbf{\alpha}$ as an example, its  posterior can be computed by the following Bayesian rules:
\begin{equation}
p(\bm{\alpha}|\textit{O})\propto p(\bm{\alpha},\textit{O})= p(\bm{\alpha})p(\bm{\textit{O}|\alpha})
\label{eq22}
\end{equation}
Unfortunately, the detail expressions of the prior distribution $p(\mathbf{\alpha})$ is unknown and too complex to be sampled.
While the variational inference\citep{blei2017variational} can be a good solution to approximate $p(\mathbf{\alpha})$.
In specific, $\bm{\alpha}$ is assumed to subject to a Gaussian distribution $q_{\phi}(\bm{\alpha})\sim\mathcal{N}(\bm{\mu},\bm{\sigma})$, where $\bm{\mu}$ and $\bm{\sigma}$ are learnable parameters. By minimizing the KL-divergence between $q_{\phi}(\bm{\alpha})$ and $p(\bm{\alpha}|\textit{O})$,the optimal $\bm{\mu}$ and $\bm{\sigma}$ can be obtained.
This training process can be formulated as maximizing the evidence lower bound (ELBO)\citep{blei2017variational}, i.e.,
\begin{equation}
\mathcal{L}^{ELBO}=E_{q_{\phi}(\bm{\alpha})}[\log p(\bm{\alpha},\textit{O})]-E_{q_{\phi}}[\log q_{\phi(\bm{\alpha})}(\bm{\alpha})]
\end{equation}
Analogously, the distribution $p(\bm{\beta}|\textit{O})$ can also be learned by a similar way.

In the action step, we aim to figure out three counterfactual distributions $p_C(C|do(M),B)$, $p_C(C|M,do(B))$ and $p_C(C|do(M),do(B))$.
Such three distributions answer three counterfactual questions, respectively, including
1) which items would a user interact with if he/she were not affected by the amplified subjective bias?
2) which items would a user like if he/she were not affected by the popularity bias?
and 3) what would the user's interaction behavior be if he is not affected by either bias?
Besides, $S_m(u,i)$ is assumed to subject to the standard normal distribution $\mathcal{N}(0,1)$ and the $do(M)$ operation is realized by sampling $S_m(u,i)$ from $\mathcal{N}(0,1)$.
Similarly, the operation $do(B)$ can be achieved by sampling $S_p(u,i)$ from $\mathcal{N}(0,1)$.

In the Prediction step, we employ the definition in Eq.\eqref{generation} to generate three types of counterfactual distribution $p_C(C|do(M),B)$, $p_C(C|M,do(B))$ and $p_C(C|do(M),do(B))$.
By selecting top-N items from these counterfactual distributions, we can construct counterfactual data $X^{counter}$, $X_m^{counter}$ and $X_p^{counter}$.
The enhanced data can be obtained by combining counterfactual data with factual data, i.e.,

\begin{equation}
 	\left\{\begin{array}{ll}
 	X^{en} = X \cup X^{counter},
 		\\
 	X_p^{en} = X_p \cup X_p^{counter},
 	\\
 	X_m^{en} = X_m \cup X_m^{counter},
 	\end{array}\right.
\end{equation}
%
%\begin{equation}
%X^{en} = X \cup X^{counter}
%\end{equation}
%\begin{equation}
%X_p^{en} = X_p \cup X_p^{counter}
%\end{equation}
%\begin{equation}
%X_m^{en} = X_m \cup X_m^{counter}
%\end{equation}
%
By retraining DB-VAE with these enhanced data, the debias difficulty due to the data sparsity can be mitigated. In this work, we set $N=100$.

%!TEX root = ./DBVAE.tex
\section{Experiments}
In this section, we conduct experiments to evaluate the effectiveness of our proposed DB-VAE.
To better guide our experimental analyses, we introduce four research questions (RQ)k, which are shown as follows:
\begin{itemize}
\item[$\bullet$] \textbf{RQ1} Does our DB-VAE framework outperform other debiasing methods?
\item[$\bullet$] \textbf{RQ2} How do different debiasing components contribute to the performance? Can counterfactual data help improve the performance of DB-VAE?
\item[$\bullet$] \textbf{RQ3} How does DB-VAE perform in the item-user groups with different data sparsities?
\item[$\bullet$] \textbf{RQ4} How does the debiasing threshold value $k$ affect the performance of DB-VAE?
\end{itemize}

{In addition, we first introduce the settings of experiment in \S \ref{sec:Experiment settings}, which includes the information about datasets, experimental setup, evaluation metrics and baseline models.
Then, we elaborate on experimental results and present some analyses in \S \ref{sec:results}, respectively answering the proposed research questions.
%from overall performance(in \S \ref{sec:Overall performance}), ablation study(in \S \ref{sec:Ablation Study}), performance on different sparsity groups(in \S \ref{sec:sparsity}) and performance with different $k$ value(in \S \ref{sec:k value})}

\subsection{Experiment settings}
\label{sec:Experiment settings}
\subsubsection{Datasets}
\label{sec:Datasets}
In this paper, we conduct experiments on three real-word recommendation datasets, including MovieLens \citep{rendle2019difficulty,zheng2016neural}, AliShop-7C \citep{ma2019learning} and Amazon-book datasets\citep{stratigi2019ratings}.
In specific,
MovieLens includes movie category information and user attribute information, which is a wildly-used dataset collected from MovieLens website.
Given the different data sizes in MovieLens, we select  ML-1M and ML-20M as our research datasets in this paper.
%We use two MovieLens datasets of different sizes, which are ML-1M and ML-20M respectively.
AliShop-7C is collected from Taobao (Alibaba's e-commerce platform).
Amazon-book is one of Amazon product datasets\citep{stratigi2019ratings}, which records how users rate different books on Amazon.
Among datasets used in our experiments, they all involve enough item features, e.g., the movie genre or the book category.
More specifically, each dataset is filtered out items with less than 5 interactions and users with less than 2 interactions to ensure data quality.
In addition, we list the information of all the datasets after pre-processing in Table \ref{table1}.
{From this table, we can find that Amazon-book is the sparsest dataset but has the largest number of categories, which may take great difficulty for model debiasing.}
\begin{table}
\center
\caption{{Statistics of datasets, where the sparsity metric is calculated by dividing the number of actual interactions by the product of the user and item numbers. {The held-out users is the number of validation/test users out of the total number of users in the first row}}}
\begin{tabular}{lcccc}

     \hline
     &ML-1M&Alishop-7c&ML-20M&Amazon-book\\
     \hline
     \#total users&6034&10668&136677&482933\\
     \#item&3533&20591&20720&367963\\
     \#interactions&575272&767497&9990682&7145464\\
     sparsity&2.70\%&0.35\%&0.35\%&0.004\%\\
     \#category&18&12&19&1600\\
     \#held-out users&800&1000&10000&40000\\
     \hline
\label{table1}
\end{tabular}
\end{table}

\subsubsection{Experimental setup}
\label{sec:setup}

To evaluate the model's ability in constructing the user representation from the obvious interaction, we take all  interactions  of a specific user as a data instance. Specifically, we split all users into training/validation/test sets and the entire click histories from the training users are employed to train models.
{The \#held-out users of Table \ref{table1} indicates the number of validating/testing users.
The 80\% of the interactions of a validation user or a test user are used as the input of different models while the remaining 20\% of the interactions are used to evaluate the model.}

For the encoder of all models (i.e., $f_{\theta}()$ in Eq.~\ref{eqencoder}), we choose a three-layer perceptions. And the decoder structure is the same as that of the encoder for the sake of symmetry.
As the dimension setting in previous work\citep{shenbin2020recvae}, we set the dimensions of the latent representation and any hidden layer to 200 and 600, respectively.
During the model training, we employ the Adam\citep{kingma2014adam} optimizer with the  batchsize of 500 users and apply a weight decay of 0.01.
And we retain the models with the best $NDCG@100$ in the validation set and evaluate them on the test set.

\subsubsection{Metrics}
\label{sec:Metrics}
We adopt two classical ranking metrics in the recommender system as the evaluating metrics, including $Recall@K$ and $NDCG@K$ \citep{nie2020large,wang2020disentangled}, which are readily appropriate to the user data never appeared in the training set\citep{wang2021deconfounded}.
By comparing the top-k predictions with the ground-truth user $u$ in the test set $\bm{X}_u^t$, $Recall@K$ and $NDCG@K$ can be obtained.

In particular, we first sort the items in descending order of the predicted likelihood scores, and then select the top-K items with the highest score to form a recommendation list $\bm{R}_u$.
Formally, $Recall@K(u)$ is defined as follows:
\begin{equation}
Recall@K(u) =\frac{\sum_{i=1}^{K} \mathbb{I} [R_u^{(i)} \in \bm{X}_u^t]}{min(K,|\bm{X}_u^t|)}
\end{equation}
where $\mathbb{I}[\cdot]$ is an indicator function that returns 1 if the condition is satisfied.
%
%The denominator returns the minimum of $K$ and $|\bm{X}_u^t|$.
And  the definition of $NDCG@K(u)$ is:
\begin{equation}
 	\left\{\begin{array}{ll}
 	NDCG@K(u)=\Big{(}\sum_{i=1}^{|\bm{X}_u^t|} \frac{1}{log(n+1)}\Big{)}^{-1}DCG@k(u),
 		\\
 	DCG@K(u)=\sum_{i=1}^{K} \frac{2^{\mathbb{I} [R_u^{(i)} \in \bm{X}_u^t]}}{log(n+1)},
 	\end{array}\right.
\end{equation}
%\begin{equation}
%DCG@K(u)=\sum_{i=1}^{K} \frac{2^{\mathbb{I} [R_u^{(i)} \in \bm{X}_u^t]}}{log(n+1)}
%\end{equation}
%\begin{equation}
%NDCG@K(u)=\Big{(}\sum_{i=1}^{|\bm{X}_u^t|} \frac{1}{log(n+1)}\Big{)}^{-1}DCG@k(u)
%\end{equation}
%
where $NDCG@K(u)$ is defined as $DCG@K(u)$ divided by its theoretically possible value\citep{shenbin2020recvae}.

Generally, $Recall@K$ is large bigger than $NDCG@K$ under the same $K$, because $Recall@K$ treats equally all items in the top-K recommendation list, while $NDCG@K$ assigns larger weights to the top-ranked items.
In this light, we select a larger $K$ for $NDCG@K$ than that for $Recall@K$ as previous work does\citep{shenbin2020recvae,liang2018variational}.
In specific, $Recall@20$ and $NDCG@100$ are adopted as evaluation metrics in our experiments.

\subsubsection{Baselines}
\label{sec:Baselines}
To verify the effectiveness of DB-VAE, we select several recent competitive recommendation models as the baselines, which can be further divided into two model groups.
%%
%We compare the performance of our proposed DB-VAE with several competitive baselines, which are proposed in recent years.
%All the baselines can be divided into two groups.
In particular, the first group is some well-designed VAE models \citep{shenbin2020recvae,liang2018variational} without the debiasing operation.
While the second group is model-agnostic debaising models \citep{wang2021deconfounded,wei2021model} which aim at eliminating either the popularity bias or the amplified subjective bias.
More specifically,
\begin{itemize}
  \item Mult-VAE\citep{liang2018variational} extends the variational autoencoder for learning implicit feedback in the top-k RS. By assuming that the user representation complies with the multivariate normal distribution, Mult-VAE introduces a different regularization parameter for the learning objective, achieving considerable performances.
  \item RecVAE\citep{shenbin2020recvae} improves Mult-VAE by several optimization methods, including a novel composite prior distribution for the latent representation, a better approach to the weight setting in the evidence lower bound and a method of alternately updating parameters.
  \item MACR\citep{wei2021model} is a model-agnostic framework that aims at eliminating the popularity bias in RS. MACR analyzes the cause-effect and introduces a multi-task learning method to answer the counterfactual question about what the ranking score would be if the model only uses item property. To fairly compare the model performance, we keep the backbone model of MACR the same as DB-VAE.
  \item DecRS\citep{wang2021deconfounded} contributes an approximation operator by a backdoor adjustment, which concentrates on eliminating the obstacles in causal reasoning theory, thereby eliminating the amplified subjective bias.
  Similar to MACR, we keep the backbone model of DecRS same as DB-VAE.
\end{itemize}

\subsection{Experimental results and analysis}
\label{sec:results}
%To answer the four research questions, We evaluate the performance of different baselines and our proposed DB-VAE on the offline datasets.
This subsection respectively answers the proposed research questions by conducting the recommendation experiments about all discussed models on the selected datasets.

\subsubsection{Overall performance.}
\label{sec:Overall performance}
For \textbf{RQ1}, we summarize the overall performance of our proposed DB-VAE as well as the selected baselines in terms of $Recall@20$ and $NDCG@100$ in Table.\ref{table2}.
%Table.~\ref{table2} presents the overall performance of baseline models and our proposed model in terms of Recall@20 and NDCG@100. The bolded result indicates the winner in that column.
Generally, for any evaluation metric on any dataset, DB-VAE consistently shows a state-of-art performance. The detailed observations are presented as follows:

\begin{table}[h]\footnotesize
\begin{center}
\caption{Performance of DB-VAE as well as the selected baselines in terms of $Recall@20$ and $NDCG@100$. The \textbf{bold font} indicates the best performer in the belonged column.}
\begin{tabular}{lcccccccc}
     \hline
    Model&\multicolumn{2}{c}{ML-1M}&\multicolumn{2}{c}{ML-20M}&\multicolumn{2}{c}{Alishop-7c}&\multicolumn{2}{c}{Amazon-book}\\
     &R@20&N@100&R@20&N@100&R@20&N@100&R@20&N@100\\
     \hline
     Mult-VAE&0.343&0.405&0.395&0.411&0.150&0.238&0.141&0.151\\
     RecVAE&0.378&0.431&0.414&0.426&0.187&0.293&0.157&0.163\\
     MACR(VAE)&0.391&0.441&0.432&0.437&0.195&0.298&0.159&0.164\\
     DecRS(VAE)&0.395&0.449&0.442&0.439&0.206&0.305&0.162&0.165\\
     DB-VAE(ours)&\bf{0.419}&\bf{0.463}&\bf{0.453}&\bf{0.461}&\bf{0.217}&\bf{0.322}&\bf{0.170}&\bf{0.177}\\
     \hline
     \label{table2}
\end{tabular}
\end{center}
\end{table}

\begin{itemize}
  \item In all cases, our disentangle debias framework boosts the performance of VAE models most obviously. Especially on Alishop-7c dataset, DB-VAE improves over 5\% compared to the best baseline model in terms of both metrics. Even the smallest one on ML-20M dataset, DB-VAE can still achieve 2.4\% improvements in terms of $Recall@20$.
  Such promising advancements over the vanilla VAE models or VAE-based debias models all demonstrate the effectiveness of DB-VAE.
  \item Clearly, VAE-based debias models (i.e., MACR, DecRS and DB-VAE) outperform vanilla VAE models (i.e., Mult-VAE, RecVAE), indicating that eliminating biases in the recommender system can be a good solution to improve the recommendation performance.
%
%  Generally speaking, the models using debias method(MACR, DecRS and DB-VAE) outperform those that do not consider eliminating bias(Mult-VAE and RecVAE), which illustrates that bias is prevalent in all datasets of the recommendation system and debias method is very important to improve the recommendation performance.
%  In the models without debias methods, RecVAE outperforms Mult-VAE a lot, which reflects that that well-designed model will greatly improve the performance.
  \item Generally, all discussed models have worse performance on the datasets with a smaller sparsity, except on ML-1M.
  This finding reflects that sparser data can be easier to make wrong predictions.
  On the other hand, the exception ML-1M has more dense data than that of ML-20M, yet models don't realize better performance on ML-1M.
  Such a contrast phenomenon may be attributed to the fact that the data volume is a more influential factor in the recommendation performance than the data sparsity.
  In other words, from the same data source, the data volume in ML-20M is 20 times of that in ML-1M, substantially making up the missing supervised signals brought by the data sparsity.
  This analysis can also be applicable to the comparison between Alishop-7c and ML-20M. In specific, despite the same data sparsity, the data volume of ML-20M is greater than that of Alishop-7c, and hence all discussed models have better performance on ML-20M than on Alishop-7c.
%%  We also notice there is an interesting phenomenon regarding the sparsity difference of data sets.
%  Compared to ML-20M dataset, ML-1M has more dense data, but the performances of all models on ML-1M are worse than that on ML-20M. This may because ML-1M collects fewer data from Movielens, which may be incomplete information. Thus the correlation between items and users can not be fully reflected, resulting in bias in recommendation. ML-20M dataset and Alishop-7c have the same sparsity, but there is a large margin between the performances of all models on these two datasets. From Table ~\ref{table2} we can see that the numbers of users and interactions are much smaller in Alishop-7c dataset than that in ML-20M, which leads to the lack of enough training samples in Alishop-7c dataset.
\end{itemize}

\subsubsection{Ablation Study}
\label{sec:Ablation Study}
Our proposed DB-VAE introduces a disentangled debias method that eliminates two different bias simultaneously.
For answering \textbf{RQ2}, we first design two variants DB-VAE models by eliminating a specific type of bias.
For example, DB-VAE(P) denotes the removal of the popularity bias in DB-VAE, where the only input of VAE is $X$ and $X_p$, and hence the user presentation in Eq.\eqref{eqsub} is transformed into $\bm{z}=\bm{z_x}-\bm{z_p}$.
%Similarly, We only keep the amplified subjective debias component in DB-VAE framework to construct DB-VAE(M).
Similarly, DB-VAE(M) represents DB-VAE without the amplified subjective debias component.
To verify the effectiveness of counterfactual data, we further introduce a DB-VAE variant combined with the counterfactual data generation module, denoted as  DB-VAE(CD).
%We also let the best baseline without debias, i.e., RecVAE take participate in comparison, so as to intuitively see the improvement of different de-bias components.
%In section ~\ref{Counterfactual}, we propose a method to generate counterfactual data in order to help train the DB-VAE model better. In this section, We define the model with counterfactual data to participate in training as DB-VAE(CD).
%By the comparison between the original DB-VAE and DB-VAE(CD). We can evaluate the proposed counterfactual data enhancement method.
Besides, we add the best vanilla VAE model (RecVAE) into the performance comparison as the lower bound of these DB-VAE variants.
With these discussed models, we summarize their performance on four different datasets in terms of  $Recall@20$ and $NDCG@100$ in Table.\ref{table3}.

\begin{table}[t]
\begin{center}
\caption{Performance of DB-VAE variants, including replacing/removing a specific component and adding a counterfactual data enhencement.The \textbf{bold font} indicates the best performer in the belonged column. R@20 and N@100 are short for $Recall@20$ and $NDCG@100$, respectively.}\footnotesize
\begin{tabular}{lcccccccc}
     \hline
     Model&\multicolumn{2}{c}{ML-1M}&\multicolumn{2}{c}{ML-20M}&\multicolumn{2}{c}{Alishop-7c}&\multicolumn{2}{c}{amazonbook}\\
     &R@20&N@100&R@20&N@100&R@20&N@100&R@20&N@100\\
     \hline
     RecVAE&0.378&0.431&0.414&0.426&0.187&0.293&0.161&0.167\\
     DB-VAE(P)&0.383&0.445&0.417&0.429&0.189&0.297&0.151&0.163\\
     DB-VAE(M)&0.352&0.423&0.421&0.433&0.185&0.295&0.157&0.165\\
     DB-VAE&\bf{0.419}&\bf{0.463}&0.453&0.461&0.217&0.322&0.170&0.177\\
     DB-VAE(CD)&0.407&0.438&\bf{0.459}&\bf{0.471}&\bf{0.219}&\bf{0.326}&\bf{0.181}&\bf{0.195}\\
     \hline
     \label{table3}
\end{tabular}
\end{center}
\end{table}

Generally speaking, disentangled debias methods show superiority over the single debias one, i.e., original DB-VAE outperforms DB-VAE(P) and DB-VAE(M) on all datasets in terms of both metrics, reflecting that the disentangled debias methods are more appropriate and effective.
In some extreme cases, the single debias variants are even worse than the model without debias components (i.e., RecVAE).
For example, DB-VAE(M) is defeated by RecVAE on ML-1M dataset in terms of both metrics and on Alishop-7c dataset in terms of $Recall@20$.
Such underperformance can be explained by the bias-coupling assumption in the introduction section, i.e., eliminating a single bias can instead cause the over-expressiveness of the other bias, further amplifying the adverse effect of
the other bias.
Furthermore, DB-VAE(P) outperforms DB-VAE(M) a lot on ML-1M with the greater data sparsity, while DB-VAE(M) takes the lead on ML-20M and amazonbook with the smaller data sparsity.
This phenomena reflect that the data-intensive recommendation system is largely affected by the popularity bias; on the other hand,  the amplified subjective bias plays a greater role in the data-sparse recommender.
%in a data-intensive situation where the elimination of popularity bias will bring great improvement to the performance.
%And vice versa, subjective biases may be more easily amplified on sparse data sets, so so de-bias operations based on the subjective bias will be more conducive to the improvement of recommendation effect.

Further, the generated counterfactual data can improve  the performance of DB-VAE on all datasets except ML-1M, especially on amazonbook where DB-VAE(CD) achieve 6\% and 10\% improvements over original DB-VAE in terms of $Recall@20$ and $NDCG@100$, respectively.
Such performance advancement indicate that counterfactual data is very appropriate to the data-sparse scenario, yet having a perverse effect on the data-intensive scenario.
Such phenomena can be attributed to the fact that counterfactual data can help supplement the lacking supervised signals when data is sparse; however, for data with sufficient interactions, the addition of counterfactual data probably change the distribution of the original data, poisoning the model performance.

\subsubsection{Performance on different sparsity groups}
\label{sec:sparsity}
To further evaluate the models' performance under different sparsity degrees (i.e., \textbf{RQ3}), we sorted the test users in ascending order according to the degree of sparsity(i.e., the number of interacted items), and then divided them equally into eight groups.
In addition, we only test debias models in this subsection, including MACR, DecRS, DB-VAE and DB-VAE(CD).
%The results are shown in the following figure:
Their group performance are summarized in Fig.\ref{group performance1}.
\begin{figure}[t]
    \centering
    \subfigbottomskip=2pt %两行子图之间的行间距
    \subfigure[ML-1M]{\includegraphics[width=0.40\hsize,height=0.35\hsize]{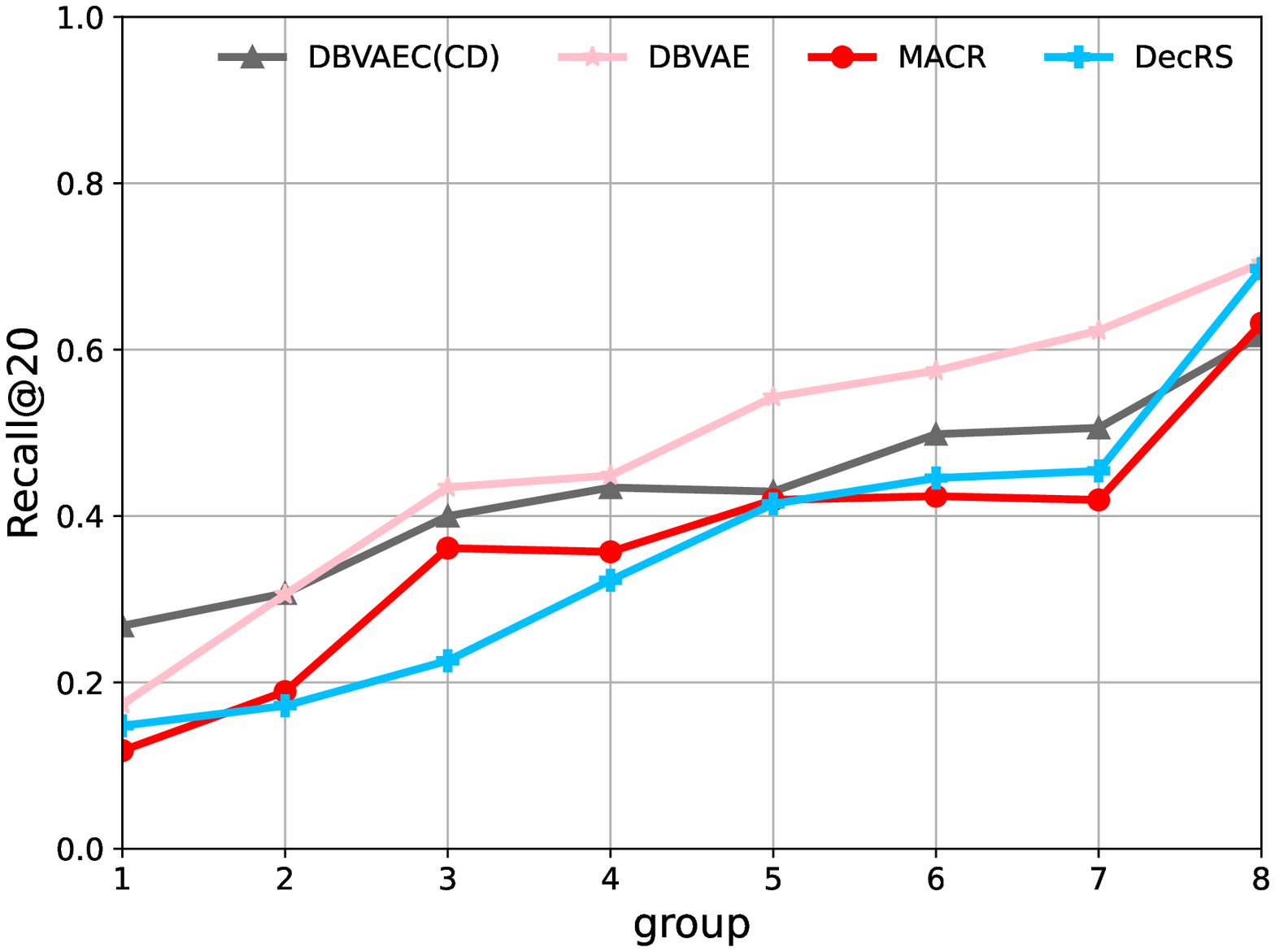}}
    \subfigure[ML-20M]{\includegraphics[width=0.40\hsize,height=0.35\hsize]{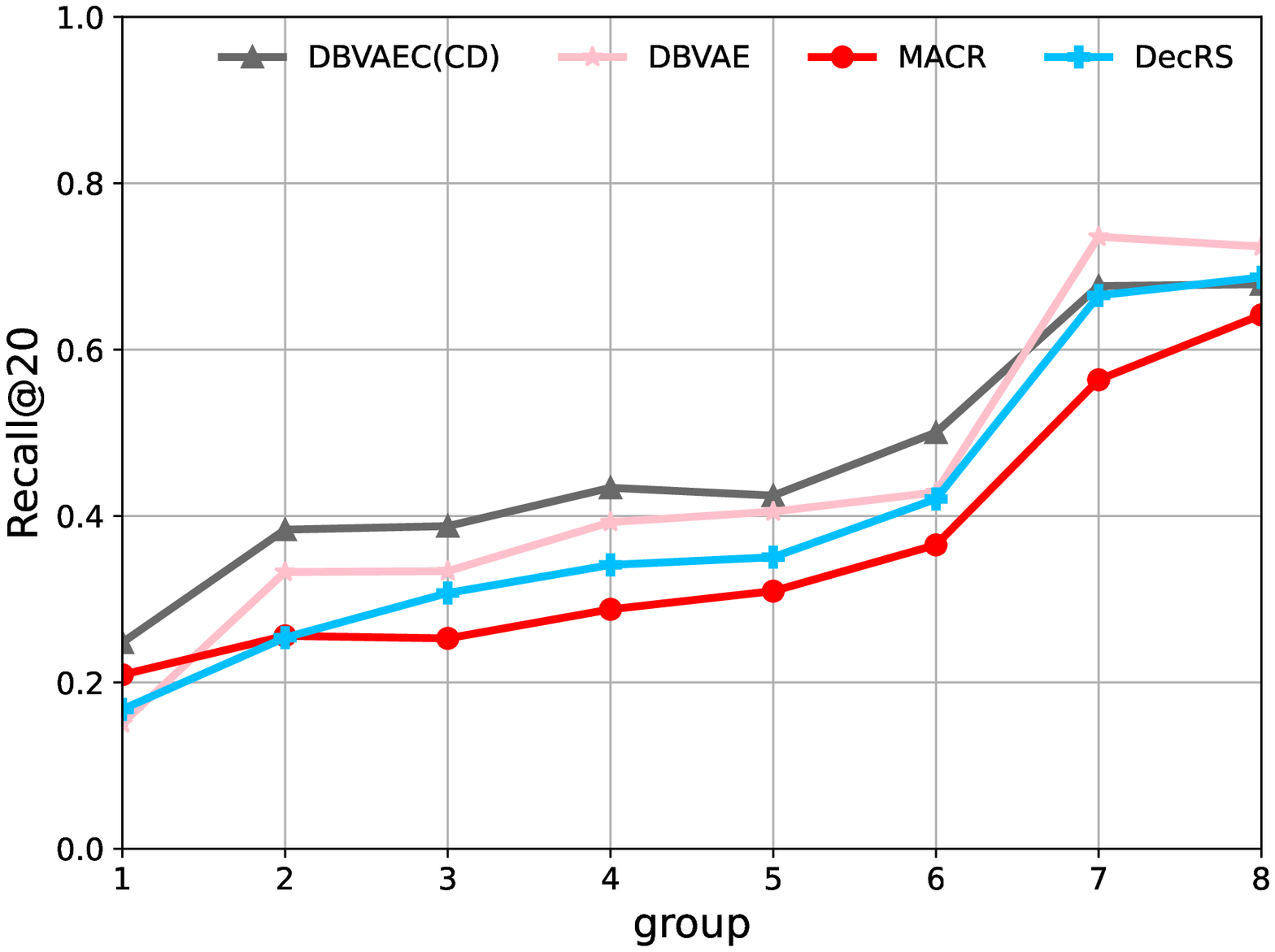}}
    \subfigure[Alishop-7c]{\includegraphics[width=0.40\hsize,height=0.35\hsize]{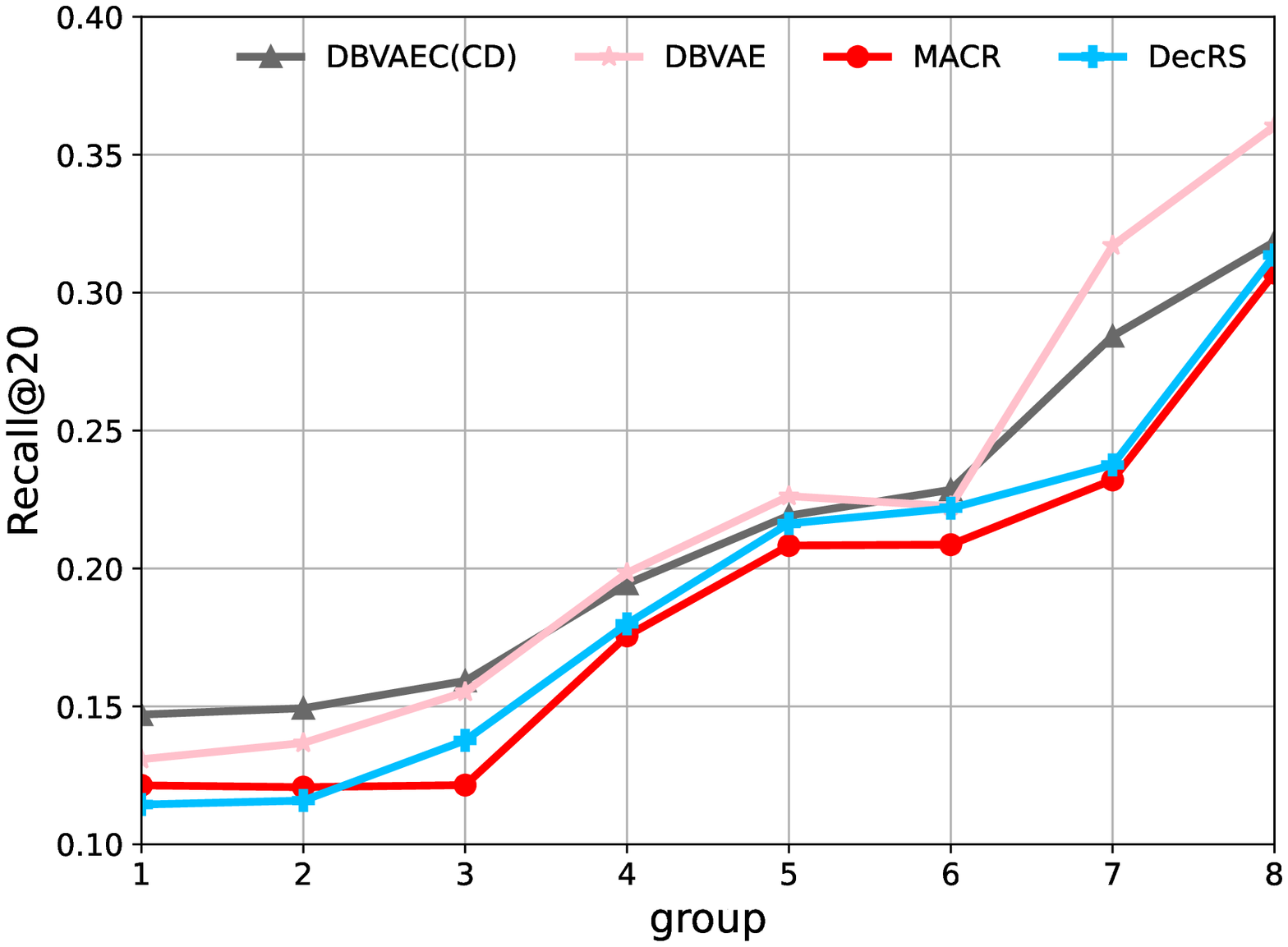}}
    \subfigure[amazonbook]{\includegraphics[width=0.40\hsize,height=0.35\hsize]{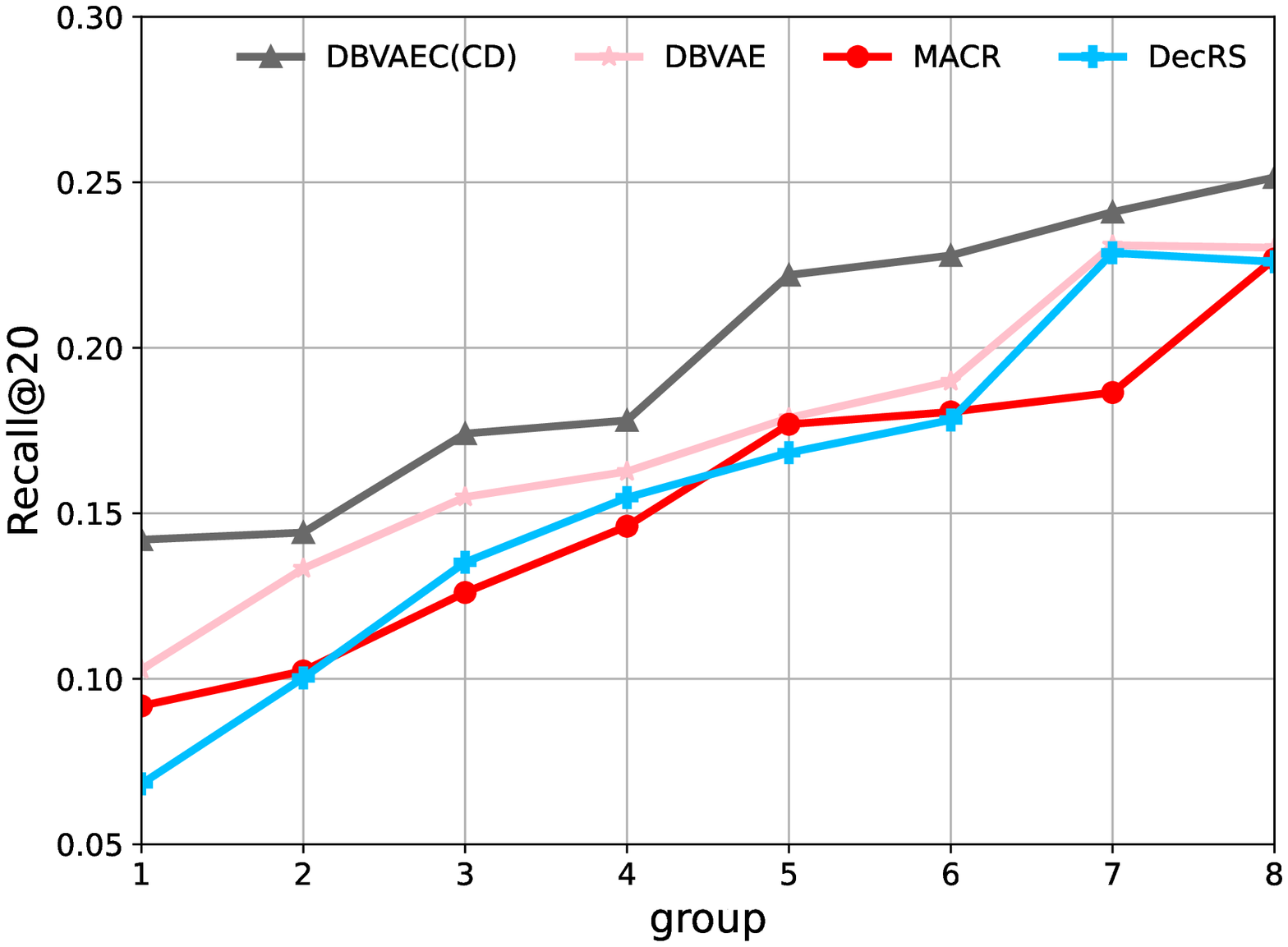}}
    \caption{Model performance on different sparsity groups in terms of Recall@20.}
    \label{group performance1}
\end{figure}
%
%The performance of all models shows an overall upward trend from group 1 to group 8 on all datasets, which indicates that data sparsity makes it very difficult to improve the performance of recommendation.
Clearly, with the increase of the sparsity value, all discussed models present an overall upward trend, indicating the data sparsity is an inherent problem that affects the recommendation performance and the smaller sparsity cause the worse model performance.
In addition, DB-VAE(CD) always achieves a greater lead in the first group on all datasets, especially on amazonbook dataset, DB-VAE(CD) outperforms the original DB-VAE by nearly 40\%.
This phenomenon further verifies that counterfactual data can indeed help to improve the effectiveness of DB-VAE in the data-sparse scenario.
On the other hand, DB-VAE becomes the best performer on ML-20M and Alishop-7c in the last two groups, demonstrating that our proposed method is most effective when there are sufficient data to support the model training.

\begin{figure}[t]
    \centering
    \subfigbottomskip=2pt %两行子图之间的行间距
    \subfigure[ML-1M]{\includegraphics[width=0.40\hsize,height=0.35\hsize]{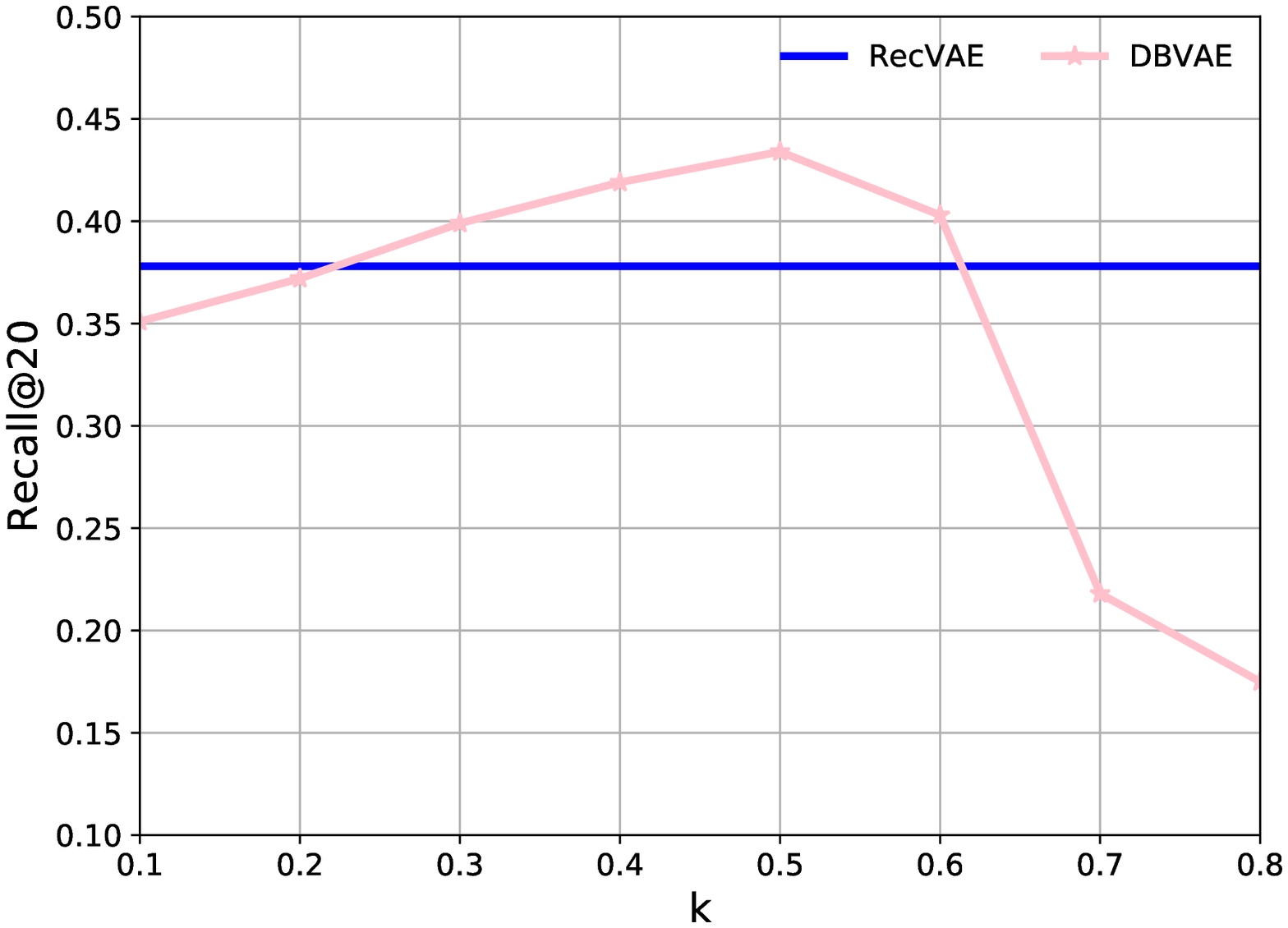}}
    \subfigure[ML-20M]{\includegraphics[width=0.40\hsize,height=0.35\hsize]{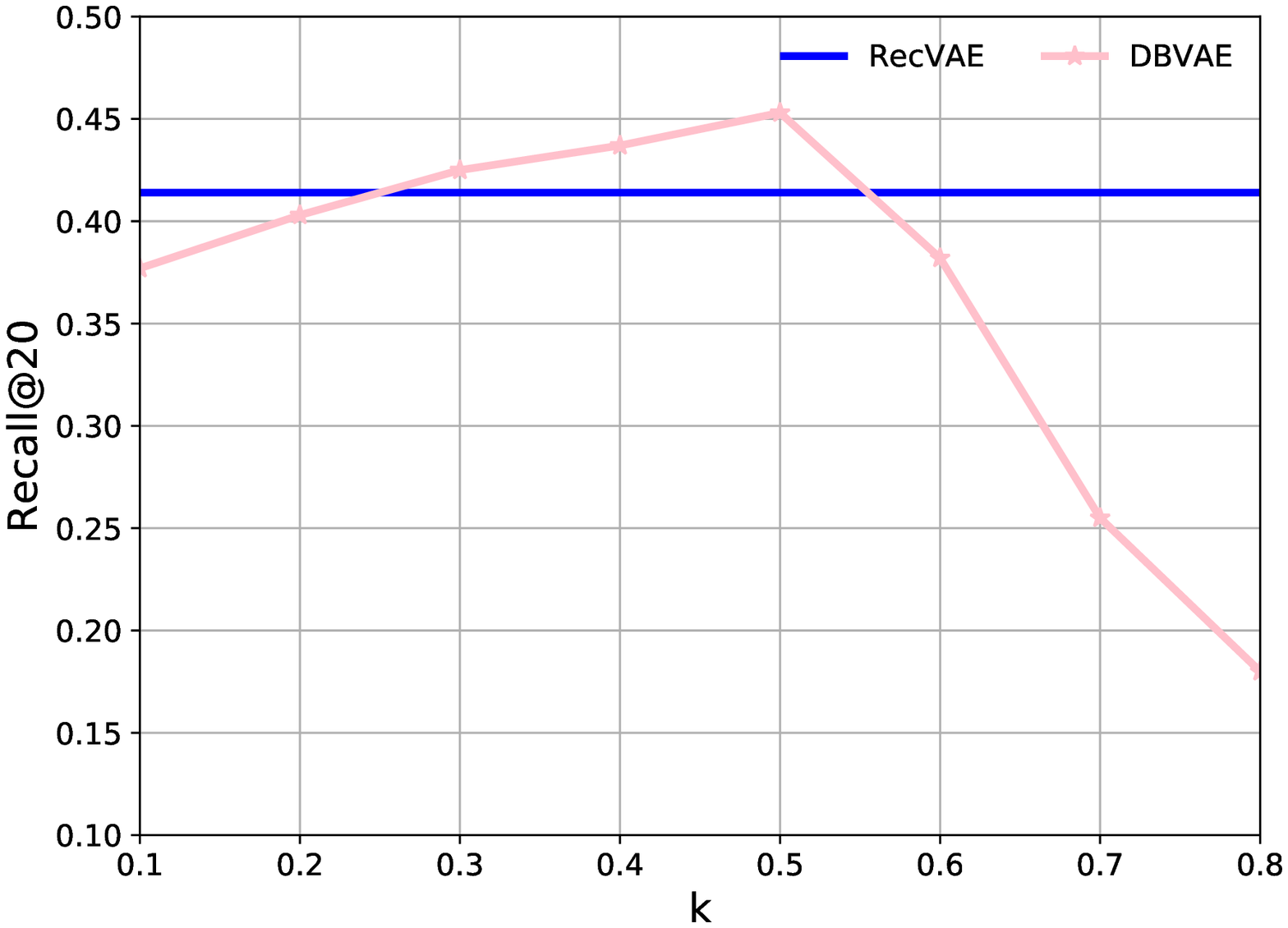}}
    \subfigure[Alishop-7c]{\includegraphics[width=0.40\hsize,height=0.35\hsize]{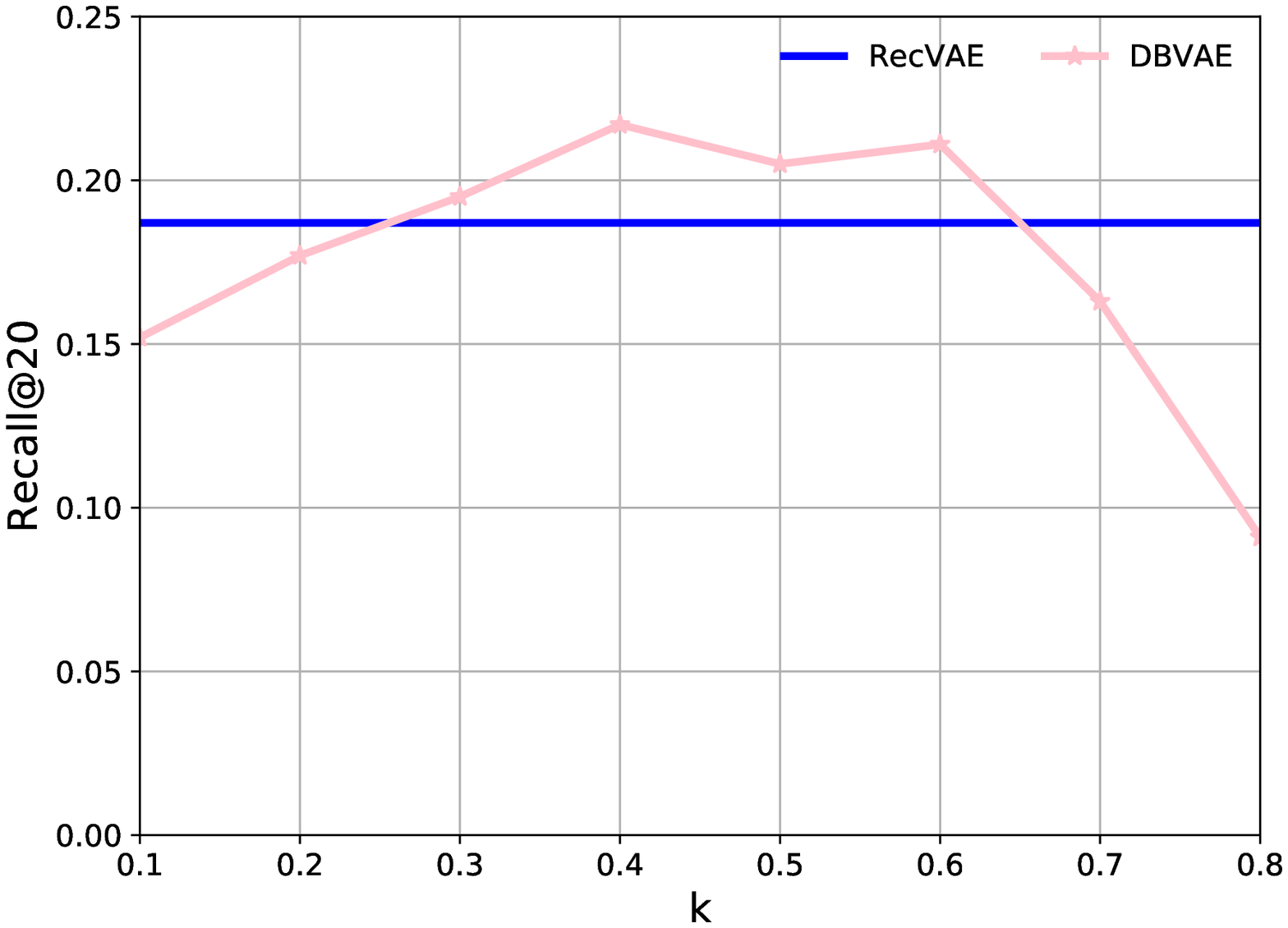}}
    \subfigure[amazonbook]{\includegraphics[width=0.40\hsize,height=0.35\hsize]{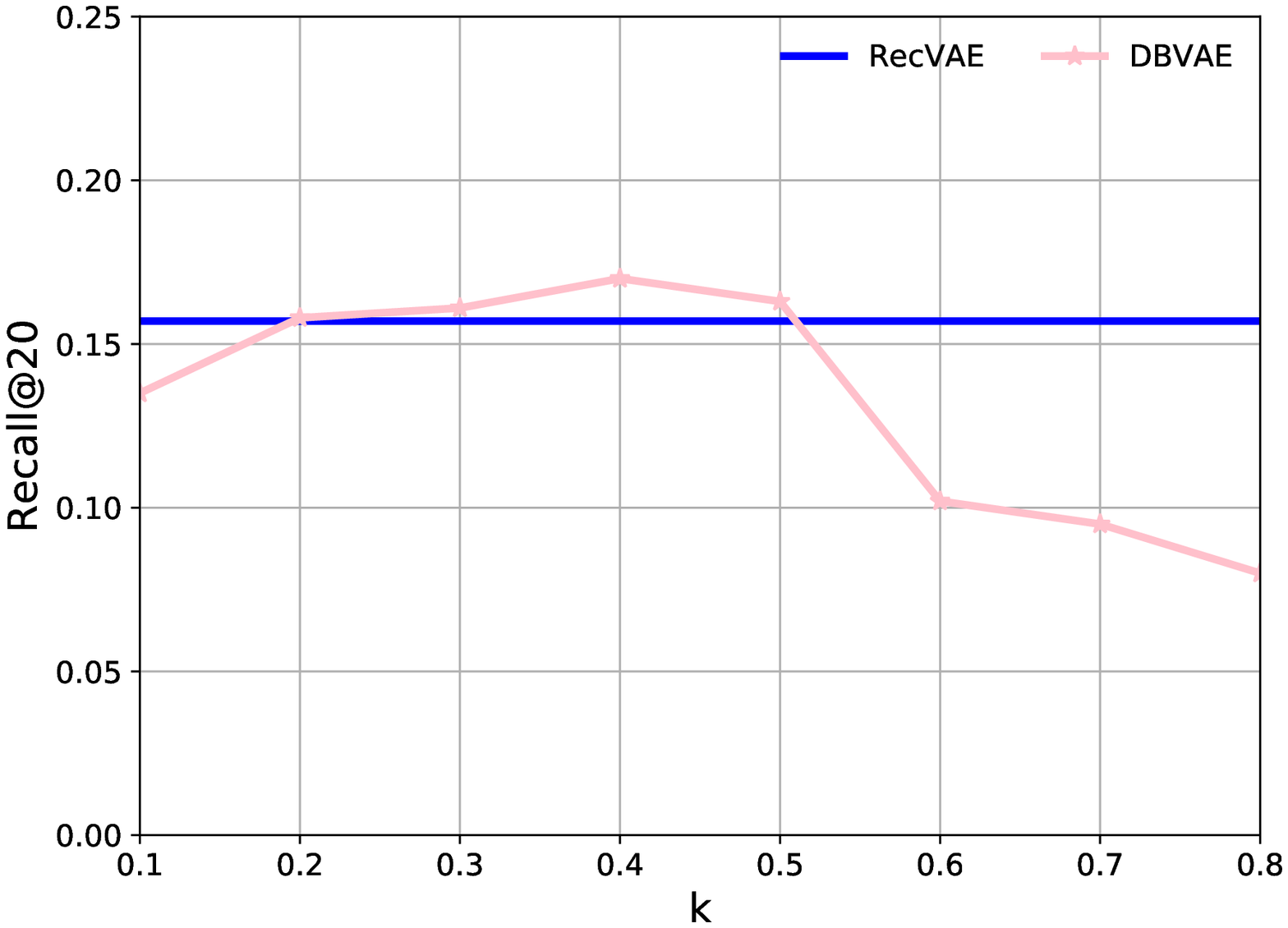}}
    \caption{Model performance on different sparsity groups in terms of Recall@20.}
    \label{group performance1}
\end{figure}

\subsubsection{How $k$ affects the performance of DB-VAE?}
\label{sec:k value}
The debias degree value $k$ in Eq.\eqref{eq4} and Eq.\eqref{eq5} controls the number of the items extracted from the user's profile. The smaller the $k$ is, the fewer items we extract to debias.
In this section, to answer \textbf{RQ4}, we evaluate how $k$ affects the performance of our proposed DB-VAE model.
For simplicity, we present the change of model performance only in terms of $Recall@20$ since the performance in terms of $NDCG@100$ has similar trends.
In addition, we also introduce the non-debias model RecVAE as the comparison benchmark to intuitively present the debias effect.
%To  intuitive view of debias effect, we select RecVAE, which does not consider debias method, as a comparison benchmark.
Hence, we summarize the performance of DB-VAE and RecVAE affected by different debias degrees $K$ in terms of $Recall@20$ in Fig.\ref{group performance2}.

%From the figure we can see that the performance of our proposed DB-VAE shows a process of first rise and then decline with the increase of $k$ on all datasets.
Generally, DB-VAE presents a consistent change pattern, i.e., increasing first and then reducing.
In addition,
DB-VAE shows the best performance on two movielens datasets when $k$ is $0.5$, while peaks at $k=0.4$ on Alishop-7c and amazonbook, indicating that different datasets have respective preferences of bias degrees.
%This indicates that different data sets have different degrees of bias, so it is necessary to find an appropriate de-bias degree to make the model perform best on the dataset.
Besides, when $k$ is small, e.g., $k=0.1$ and $k=0.2$,
% although DB-VAE does not achieve its own best performance, the performance is only a little worse than RecVAE.
DB-VAE exhibits a little behind RecVAE in the model performance.
As $k$ becomes bigger until the peak, the performance of DB-VAE constantly gets improved, exceeding RecVAE in some $k$ values.
However, when $k$ is greater than $0.6$, DB-VAE presents a steep performance degeneration, creating a huge gap against RecVAE.
These findings demonstrate that on one hand, slight debias cannot improve the model performance; on the other hand, excessive debias causes more damage to the model performance.
%
%the model performance on the data set cannot be improved with slight debias, excessive debias will cause more damage to the model performance.
Therefore, locating the appropriate debias degree becomes an important issue in future recommendation system research.

\section{Conclusion}

In this work, we explain the two most common biases in the recommender system, popularity bias and amplified subjective bias. To alleviate these biases, we propose a disentangled debias variational auto-encoder framework, which overcomes the shortcomings of other debias methods that are single and have no de-biased supervisory signals. In the process of debias, we make use of the relevant theory of causal inference, which helps us find the user interaction that may be biased. Extensive experiments validate that our proposed DB-VAE outperforms other debias methods. Also, we design a data enhancement method to help model training when data is sparse.

In future work, we will try to find the relationship between other types of biases and construct a unified and effective debias framework.

\bibliography{mybibfile}

\end{document}